%% file: ms.tex
\documentclass[11pt]{article}

\usepackage[11pt]{moresize}
\usepackage{times}
\usepackage{authblk}
\usepackage[pdftex]{graphicx}
\usepackage{xcolor}
\usepackage{color}
\usepackage{colortbl}
\usepackage{xspace}
\usepackage{amsmath,amscd,amssymb}
\usepackage{subfigure}
\usepackage{boxedminipage}
\usepackage{comment}
\usepackage{url}
\usepackage{enumitem}
\usepackage{wrapfig}
\usepackage{booktabs} 
\usepackage{mathrsfs}
\usepackage{balance}
\usepackage{algorithmic}
\usepackage[algoruled]{algorithm2e}
\usepackage{cite,alltt}
\usepackage{float}
\usepackage[rflt]{floatflt}
\usepackage{array}
\usepackage[labelfont=bf]{caption}
\usepackage{url}

\usepackage{breakurl}

\makeatletter
\newcommand\email[2][]%
{\newaffiltrue\let\AB@blk@and\AB@pand
	\if\relax#1\relax\def\AB@note{\AB@thenote}\else\def\AB@note{\relax}%
	\setcounter{Maxaffil}{0}\fi
	\begingroup
	\let\protect\@unexpandable@protect
	\def\thanks{\protect\thanks}\def\footnote{\protect\footnote}%
	\@temptokena=\expandafter{\AB@authors}%
	{\def\\{\protect\\\protect\Affilfont}\xdef\AB@temp{#2}}%
	\xdef\AB@authors{\the\@temptokena\AB@las\AB@au@str
		\protect\\[\affilsep]\protect\Affilfont\AB@temp}%
	\gdef\AB@las{}\gdef\AB@au@str{}%
	{\def\\{, \ignorespaces}\xdef\AB@temp{#2}}%
	\@temptokena=\expandafter{\AB@affillist}%
	\xdef\AB@affillist{\the\@temptokena \AB@affilsep
		\AB@affilnote{}\protect\Affilfont\AB@temp}%
	\endgroup
	\let\AB@affilsep\AB@affilsepx
}
\makeatother

\AtBeginDocument{%
	\providecommand\BibTeX{{%
			\normalfont B\kern-0.5em{\scshape i\kern-0.25em b}\kern-0.8em\TeX}}}

\input{macros}

\begin{document}

\title{GPU-Powered Spatial Database Engine for Commodity Hardware: Extended Version}

\author[1]{Harish Doraiswamy}
\author[2]{Juliana Freire}

\affil[1]{Microsoft Research India}
\email{\textit{harish.doraiswamy@microsoft.com}}

\affil[2]{New York University}
\email{\textit{juliana.freire@nyu.edu}}
\date{}

\maketitle
\input{abstract}

\input{intro}
\input{background}
\input{system}

\input{canvas}

\input{operators}

\input{exp}

\input{discussion}

\input{related}

\input{conc}

\subsection*{Acknowledgements} 
This work was partially supported by the DARPA D3M program and NSF award IIS-2106888.  Any opinions, findings, and conclusions or recommendations expressed in this material are those of the authors and do not necessarily reflect the views of NSF and DARPA.

\bibliographystyle{abbrv}
\bibliography{paper}  

\end{document}

%% file: macros.tex
\newcommand{\fig}     [1] {Fig.~\fignum{#1}}
\newcommand{\figs}    [1] {Figs.~\fignum{#1}}
\newcommand{\fignum}  [1] {\ref{#1}}

\newcommand{\ie}{{i.e.,}\xspace}
\newcommand{\eg}{{e.g.,}\xspace}
\newcommand{\dataset}{data set\xspace}

\newcommand{\datasets}{data sets\xspace}
\newcommand{\Datasets}{Data sets\xspace}
\newcommand{\DataSets}{Data Sets\xspace}

\newcommand{\myparagraph}[1]{\vspace{0.1in}\noindent \textbf{#1}}
\newcommand{\myparagraphem}[1]{\noindent \emph{#1}}

\newcommand{\spade}{{\tt Spade}\xspace}
\newcommand{\spadebf}{{\tt\bf Spade}\xspace}

\newcommand{\knn}{{$k$NN}\xspace}
\newcommand{\io}{{I/O}\xspace}

{\begin{list}{$\bullet$}{%
			\setlength{\labelsep}{2pt}\setlength{\leftmargin}{0pt}%
			\setlength{\labelwidth}{0pt}%
			\setlength{\listparindent}{0pt}}}
	{\end{list}}
\newcommand{\denselist}{\itemsep 0pt\parsep=1pt\partopsep 0pt}

\def\C{\mathbb{C}}

%% file: abstract.tex
\begin{abstract}
Given the massive growth in the volume of spatial data, there is a
great need for systems that can efficiently evaluate
spatial queries over large data sets.
These queries are notoriously expensive using traditional database solutions.
While faster response times can be attained through powerful clusters
or servers with large main-memory, these options, due to cost and complexity, are out of reach
to many data scientists and analysts making up the long tail. 

Graphics Processing Units (GPUs), which are now widely available even in
commodity desktops and laptops, provide a cost-effective alternative
to support high-performance computing, opening up new opportunities
to the efficient evaluation of spatial queries.
While GPU-based approaches proposed in the literature have shown great
improvements in performance, they are tied to specific GPU hardware
and only handle specific queries over fixed geometry types.

In this paper we present SPADE, a GPU-powered spatial database engine that
supports a rich set of spatial queries.  
We discuss the challenges involved in attaining efficient query
evaluation over large datasets as well as
portability across different GPU hardware, and how these are addressed
in SPADE.
We performed a detailed experimental evaluation to assess the
effectiveness of the system for wide range of queries and datasets,
and report results which show that SPADE is scalable and able to
handle data larger than main-memory, and its performance on a laptop
is on par with that other systems that require clusters or large-memory servers.
\end{abstract}

%% file: intro.tex
\section{Introduction}
\label{sec:intro}

The widespread use of GPS-based sensors, IoT devices, phones and social
media has resulted in an explosion of spatial data
that is being collected and stored. Increasingly, data scientists and analysts
are resorting to interactive analyses and visual tools to obtain
insights from these data (e.g., \cite{arcgis,taxivis,ferreira@vast2015,AndrienkoSurvey,tableau}).
However, executing queries involving complex geometric constraints
over large spatial \datasets is time consuming, and this greatly hampers
users' ability to interactively explore data. 
Commonly-used approaches, such as the spatial extensions available in
existing relational database systems, even on powerful
hardware, take from several minutes to run a
single spatial selection query~\cite{STIG2016} to several hours for
more complex queries such as joins~\cite{transformers}.
While faster response times can be attained through powerful clusters
and large memory servers, these options, due to cost and complexity,
are often out of reach for analysts that make up the long tail; they
are forced to work with small subsets of the data that can be handled
in less powerful desktops and laptops. This raises an important
question: can we support scalable spatial analytics for users in the long tail?

Modern graphics processing units~(GPU) exhibit impressive
computational power and provide a cost-effective alternative to
support high-performance computing. For example, the Intel UHD GPU
present in mid-range laptops reaches up to 0.5 TFLOPS, while the
previous generation Nvidia GTX 1070 MaxQ mobile GPU reaches over 5
TFLOPS. Current-generation desktop and server GPUs are even faster,
attaining speeds as high as 36~TFLOPS.
Their wide availability, even in lower-end laptops, open up new
opportunities to democratize large-scale spatial analytics.
Not surprisingly, methods have been proposed to evaluate different types
of spatial queries on the
GPU~\cite{STIG2016,Zhang2012a,Zhang2012b,Zhang2012d,Bustos2006,rasterjoin}. However, 
their adoption has not expanded beyond select research projects.  We
posit that this is due to two main reasons.
First, no single system or library can both harness the power offered by GPUs 
and support the variety of queries commonly used in spatial analysis.
Second, almost all current approaches use CUDA~\cite{cuda} and only
work on Nvidia GPUs which are not always available in commodity
hardware.

One possible approach would be to design a full-fledged GPU-based
spatial query engine that combines existing approaches. But such a
strategy quickly becomes unwieldy because existing methods target
specific queries (e.g., selection over point data, or aggregation over
polygons) and use custom data structures/indexes for each query type.
This makes it difficult to reuse the data structures across different
queries.
Moreover, many of these approaches port CPU-based algorithms to
the GPU. While general-purpose computing capabilities of GPUs have
improved in recent times, traditional algorithmic constructs still
cannot make optimal use of the parallelization provided by them, often
leading to ineffective use of GPU capabilities.

\myparagraph{The \spadebf Query Engine.}
To overcome these challenges, we propose \spade, a GPU-powered spatial
query engine.  By adopting the \textit{canvas data model and
  GPU-friendly algebra}~\cite{gpu-algebra}, \spade is able to support
a rich set of query types. In addition, since the algebraic operators
are adapted from common computer graphics operations for which GPUs
are optimized, it is possible to harness the compute power provided by
the hardware.  \spade uses the computer graphics pipeline to implement
the operators, and thus, it is portable and can be run on any GPU
hardware. 

There are several challenges involved in implementing the GPU-friendly
algebra and data model to efficiently handle large spatial \datasets
on commodity hardware. First, the system must support datasets that do
not fit in main memory.  While the implementation of some of the
algebra operators is
straightforward, for others a na\"ive implementation leads to
performance bottlenecks and the inability to handle large data due to
memory limitations of the GPU. In addition, these operators involve
costly geometric tests whose computational complexity is
polynomial to the size of the geometries, and thus can negatively
impact query execution time.

To design \spade, we adopted a computer-graphics perspective to
evaluate spatial queries. 
We devised a methodology to implement the GPU-friendly
operators~\cite{gpu-algebra} using the modern graphics
pipeline~\cite{redbook}, which enables both portability and
efficiency.
We show that, by using the GPU-based operators to efficiently query
spatial indexes, \spade is able to handle large data sets that do not fit
in memory.
We also propose two canvas-specific indexes that both speed up
geometry intersections and improve GPU occupancy.

\spade is designed to seamlessly support linkage to relational data
and to be easily integrated with relational database systems.  This is
crucial since analytics over spatial data often requires a combination
of relational and spatial queries.
\spade also includes a query optimizer that specifies the order of operations
for a given query plan and chooses the appropriate index and/or implementation
for the different operators.

To assess the performance of \spade and its ability to handle
large-scale analytics on commodity desktops and laptops, we perform an
extensive evaluation using a wide range of queries and real \datasets
having as many as 100~million polygons and 2~billion points.
The results show that \spade is scalable and efficiently handles data
that is larger than main memory and that its performance is comparable
to that of cluster-based and large memory solutions.
Specifically, \spade running on a laptop equipped with a
Nvidia 1070 MaxQ GPU performs on par with both 
GeoSpark~\cite{geospark} running on a 17 node cluster,
as well as in-memory spatial libraries~\cite{google-s2} running on 
large main-memory servers. 
We also discuss insights uncovered during the evaluation into how to
further improve \spade as well as how to combine it with the existing
solutions.

\myparagraph{Contributions.}
We introduce \spade which, to the best of our knowledge, is the first
GPU-powered spatial database engine that
supports the evaluation of a rich set of spatial queries over
different GPU hardware. Given its ability to efficiently handle complex queries over
big data on commodity hardware, \spade has the potential to make large-scale
spatial data analysis widely accessible.
To summarize: 

\begin{itemize}[leftmargin=10pt]\denselist
\item We demonstrate how the canvas model and the spatial
  algebra~\cite{gpu-algebra} can be realized using the computer
  graphics pipeline to attain efficiency, accuracy, portability, and
  scalability on commodity hardware.
\item We propose two indexes designed to support the canvas model: the
boundary index that enables constant time geometry intersection
tests, and the layer index that speeds up spatial join queries by
improving the GPU occupancy.
\item We describe the architecture and implementation of the \spade
  system and show how it can be integrated with existing relational
  systems, how existing disk-based spatial indexes can also be queried
  using the GPU operators, and how query execution can be optimized.
\item We present results of an extensive experimental evaluation of
  \spade which demonstrate its efficiency. 
\end{itemize}

%% file: background.tex
\section{Background}
\label{sec:bkgd}

To provide context
we first give a brief
informal overview of the canvas data model and
GPU-friendly spatial algebra~\cite{gpu-algebra}. 
We refer the reader to
\cite{gpu-algebra,gpu-algebra-full} for details.
We then describe the shader-based computer
graphics pipeline~\cite{redbook} that is used to implement \spade.

\subsection{Spatial Data Model and Algebra} \label{section:model-algebra}

The canvas data model and GPU-friendly algebra~\cite{gpu-algebra}
were designed to exploit GPUs for executing spatial queries. 
Intuitively, the \textit{canvas} data model~\cite{gpu-algebra} 
can be thought of as a ``drawing" of the geometry, and 
is represented as a collection of images.
However, unlike traditional images, each pixel of a canvas model image
is designed to store the necessary metadata to both identify the geometry it represents
as well as to support query execution.
This model provides a uniform representation for spatial objects -- 
not only points, lines, and polygons, but also any combination of these primitives.

The main advantage of the above representation is that
it naturally allows the parallel execution of the GPU-friendly operators
defined by the algebra. 
In particular, the algebra consists of five fundamental operators, inspired by common
computer graphics operations. Figure~\ref{fig:operators-summary} illustrates the
operators implemented in \spade.

\myparagraphem{1. Geometric Transform:}
moves geometric objects in space.

\myparagraphem{2. Value Transform:}
modifies a geometric object's properties.

\myparagraphem{3. Mask:}
filters regions 
based on a given mask condition.

\myparagraphem{4. Blend:}
merges two canvases into one.

\myparagraphem{5. Dissect:}
splits a canvas into a collection of non-empty canvases.

In addition to these, the algebra also includes derived operators that 
are compositions of one or more fundamental operators. These
include the \textit{Multiway Blend} operator, that applies several blends to merge multiple
canvases, and the \textit{Map} operator that is a composition of a dissect followed by
a geometric transform. 
Spatial queries are realized by composing one or more of the above operators.
For example, consider the spatial selection query that selects all 
geometric objects intersecting a polygonal constraint.
This is realized by first blending the canvas corresponding to a geometric object with the canvas
representing the query polygon, and then applying the mask operator that 
removes objects outside the query polygon.
The first step creates an image that merges the query polygon and the input geometry into a single canvas,
and the second step removes any geometry that is present outside the query polygon. The non-empty canvases that remain
after the above two operations correspond to the query results. Note that the above composition of operators
work on any geometric primitive.
As described in Section~\ref{sec:query-execution}, other spatial queries can be realized 
in a similar fashion.

\subsection{The Computer Graphics Pipeline}

Graphics intensive applications, such as games, involve rendering
complex scenes that continuously change based on end-user 
interactions.
These scenes are rendered as a collection of \textit{frames}, where each
frame contains the geometric objects as seen from a camera at a given time stamp.
To obtain smooth transitioning between the frames, it is critical to have
a high frame rate, \ie a large number of frames rendered per second.
GPUs are designed to speedup precisely these operations which are executed
through the computer graphics pipeline -- a high-level interface 
that allows the use of the underlying GPU hardware, while the GPU drivers
handle low-level tasks such as optimizing for the GPU architecture 
(\eg handling rasterization, threads). 

OpenGL~\cite{redbook} (cross platform), Direct3D~\cite{directx} (Microsoft Windows), metal~\cite{metal} (OSX),
and the more recent Vulkan~\cite{vulkan} (cross platform) are 
some of the popular application programming interfaces (API) that 
support the computer graphics pipeline.
This pipeline (also known as the \textit{shader pipeline}) is divided
into a set of stages
that can be
customized by the developer through the use of \textit{shaders}.

\myparagraph{Vertex Shader.}
The first stage of the 
pipeline is used to process the set of individual vertices that
form the \textit{geometric primitives} to be rendered.
It supports three types of geometric primitives---\textit{points}, 
\textit{lines} and \textit{triangles}
(polygons are rendered as a collection of triangles).
This stage is used to convert vertex coordinates into
a common \textit{screen space} (the coordinate system w.r.t. the camera),
known as \textit{model-view-projection}.

\myparagraph{Custom Geometry Creation.}
This is an optional stage which 
is used to create custom geometries. 
Either the \textit{tessellation shader} or 
the \textit{geometry shader} can be used for this purpose.
This stage takes as input the vertices output from the
vertex shader together with optional user-defined parameters,
and outputs zero or more new geometric primitives.

\myparagraph{Vertex Post Processing.}
This stage is handled by the GPU (driver) and is
used for \textit{clipping} and \textit{rasterization}.
\textit{Clipping} is the process where primitives 
outside the \textit{viewport} (the space corresponding to the region 
visible from the camera) are removed, while those (lines and triangles) 
that are partially outside are cropped resulting in
a new set of primitives that are fully contained within
the viewport.

\textit{Rasterization} is the process that converts each primitive 
within the viewport into a collection of 
\textit{fragments}.
Here, a \textit{fragment} can be considered as the \textit{modifiable data} corresponding to
a pixel. The fragment size therefore depends on the \textit{resolution} 
(width and height in terms of the number of pixels that form the screen space).
Given the crucial part it plays in the graphics pipeline, GPU hardware vendors 
optimize parallel rasterization by directly mapping computational concepts 
to the internal layout of the GPU.
Since all the primitives are simple (lines or triangles), identifying
fragments that form a given primitive is amenable to hardware implementation.

\myparagraph{Fragment (or Pixel) Shader.}
This stage allows custom processing for each fragment that is generated
by the rasterizer. It is typically used to compute and 
set the ``color" and ``depth" (distance from the camera)
for the fragment. 
Depending on the required functionality, it can also be used for other
purposes (e.g., to discard fragments, write to additional output buffers).

\myparagraph{Post Fragment Processing.}
The final stage of the pipeline processes the individual fragments output from the 
fragment shader to generate the pixels of the image corresponding to the scene being rendered.
It is responsible for performing actions such as 
\textit{alpha blending} (when objects have translucent material, then the colors 
of obstructing objects are blended together based on a blend
function).

\myparagraph{Virtual Screen.}
Graphics APIs also allow results to be output to a ``virtual" screen,
which is represented using a \textit{frame buffer} (also
called \textit{frame buffer object} or \textit{FBO}).
Thus, rendering can now be performed without the need of
a physical monitor.
Even though the resolutions supported by
existing monitors are limited, current generation graphics hardware
supports frame buffers with resolutions as large as 32K~$\times$~32K.

Each pixel of this FBO is represented by 4 values $[r,g,b,a]$, 
corresponding to the red, blue, green, and alpha color channels.
Each FBO is associated with a \textit{texture} that stores the actual image. 

%% file: system.tex
\section{SPADE: System Overview}
\label{sec:overview}

\begin{figure}[t]
	\centering
	\includegraphics[width=0.7\linewidth]{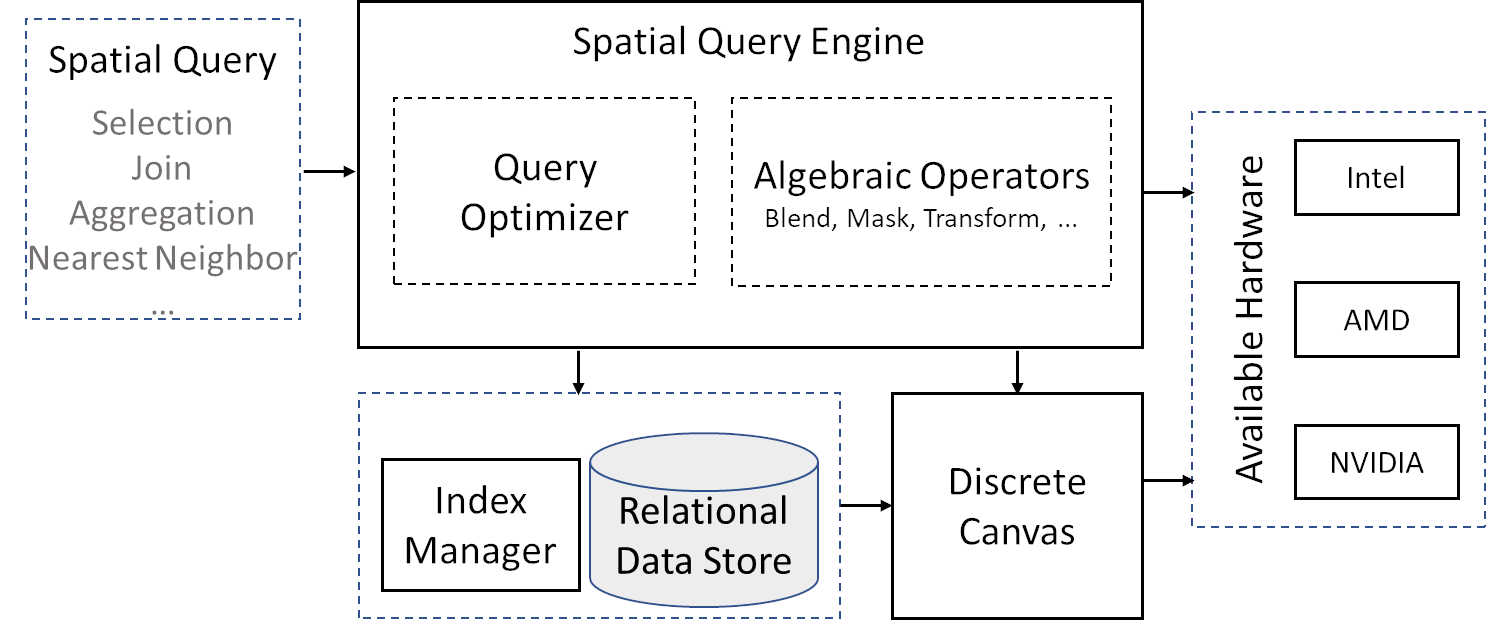}
	\caption{\spade system architecture.}
	\label{fig:spade-design}
\end{figure}

The architecture of \spade, shown in \fig{fig:spade-design}, consists
of four main components which we describe below.
Note that since \spade is implemented using the computer graphics pipeline, 
it can run on any GPU hardware. 

\myparagraph{Relational Data Store.}
All data, indexes, and meta-data used by \spade are stored as
relational tables.
For our initial implementation, we use the embedded column store
MonetDBLite~\cite{monetdblite-paper} which provides a simple C API
to load and store data using SQL.
However, since the data is accessed using SQL, it is straightforward to integrate 
\spade with other database systems.

\myparagraph{Discrete Canvas Creation.}
This component is responsible for creating canvases -- the data used by the spatial operators.
In Section~\ref{sec:representation}, we outline how a discrete canvas can be 
created using the shader pipeline, and describe how to ensure accurate queries
in spite of using rasterization to generate this canvas.
\spade supports \datasets comprising of common geometric primitives---points, 
lines, and polygons\footnote{wlog., lines and polygons are used to denote polylines and multi-polygons.}.

\myparagraph{Spatial Query Engine.}
The spatial query engine, described in Section~\ref{sec:query-engine}, forms the
core of \spade and is responsible for planning, optimization, and execution of
spatial queries.

\myparagraph{Index Manager.}
\spade maintains two types of indexes that are used during the execution of spatial queries:

\myparagraphem{1. Canvas-specific indexes}
speed up spatial intersection tests and reduce the number of
canvases that are created respectively, thus making spatial operations
on the GPU more efficient (see Section~\ref{sec:canvas-index}).

\myparagraphem{2. The disk-based index}
supports out-of-core queries -- the current version of \spade stores the 
underlying spatial data using a clustered
grid index. 
However, as discussed in Section~\ref{sec:discussions}, other indexing strategies can be easily
integrated with \spade.

%% file: canvas.tex
\section{Discrete Canvas}
\label{sec:representation}

The canvas, when rasterized by the GPU, can lead to inaccurate query results.
To overcome this limitation, we formalize the notion of discrete canvas in Section~\ref{sec:texture-representation}
and describe its creation using the rasterization-based pipeline in 
Section~\ref{sec:canvas-creation}.
Finally, in Section~\ref{sec:canvas-index}, we introduce canvas-specific indexes
that we designed to speedup query processing.

\subsection{Rasterized Canvas Representation}  \label{sec:texture-representation}
Using the rasterization-based pipeline, the Euclidean space
is divided into a set of pixels. Thus, a trivial implementation 
of the canvas would have each pixel map to the necessary metadata.
This metadata is defined as a triple of triples (or a $3\times3$ matrix)~\cite{gpu-algebra}, 
where each triple stores information corresponding to 
the three primitive types (point, line, polygon) respectively, 
that makeup a geometric object.
However, this would result in inaccuracies due to the discretization caused 
by rasterization: since a pixel need not lie entirely within a
geometry, this would induce errors for pixels that partially intersect the geometry.

We therefore extend the formal model from~\cite{gpu-algebra} to define a \textit{discrete canvas}
that uses a \textit{triple of 4-tuples}(or a $3\times4$ matrix). 
Here, for each 4-tuple $(v_0,v_1,v_2,v_b)$, $v_0,v_1$, and $v_2$ are the same as in
the original definition, and $v_b$ stores the necessary boundary data -- a pointer to the 
\textit{boundary index} that is used to obtain accurate query results 
(see Section~\ref{sec:canvas-index}).
Intuitively, the discrete canvas can be thought of as a ``localized" grid index corresponding to 
each geometric object.
As mentioned earlier, each pixel of the texture associated with an FBO can
store 4 values corresponding to the 4 color channels---this 4-tuple directly maps to the color channels of a pixel.
Thus, to represent the
canvas on the GPU, we use 3 textures, one for each primitive type.

\subsection{Canvas Creation}
\label{sec:canvas-creation}
\spade uses canvases to represent both data and query constraints.
In our implementation, instead of storing all canvases associated with a
\dataset, we create them on the fly as needed.
The advantages of this approach are twofold.  First, the data can be stored in
the database in its vector format, thus reducing the space overhead.
Second, depending on the query constraints, it is
possible to realize the canvas for only a subset/sub-region of the
data, often making it more efficient to render than load a pre-stored
canvas.
This is because the memory overhead of the serialized canvas
is typically much higher, making it more costly to transfer the data from
CPU to GPU as compared to data stored in a vector format.
In what follows, we describe how canvases are created for different types of geometries.

\myparagraph{Canvases for Traditional Primitives.}
We use the shader pipeline consisting of the vertex and fragment shaders to create canvases 
corresponding to traditional geometric primitives, \ie points, lines, and polygons.
Polygons are first decomposed into a set of triangles. We use the ear
clipping based approach implemented in the Earcut.hpp
library~\cite{earcut.hpp} for this purpose.

The vertex shader takes as input the coordinates of the vertices that
make up the geometry (end points of the lines in case of lines, and
vertices of the triangles in case of polygons) together with a
unique identifier of the geometric object they are part of. 
A transformation matrix is also passed to the vertex shader, that converts the vertex 
coordinates that fall within the valid query region
to the range $[-1,-1] \times [-1,-1]$
(Section~\ref{sec:query-engine} describes how this region is defined).
Any additional coordinate system projection 
(\eg converting from degree based EPSG:4326 coordinate system to the meter-based
EPSG:3857 coordinate system) is also performed in the vertex shader.

The fragment shader then creates the canvas by writing the object identifier to the texture attachment in the FBO
corresponding to the primitive being processed.
Note that the fragments processed in the fragment shader correspond only to the primitives
within the query region due to the clipping performed by the vertex post processing stage.
For lines and polygons, in addition to the object identifier, the pointer to the
boundary index is also written to the texture. For polygons, this process is
done in two passes. The first pass renders the triangles, while the second pass
renders the lines forming the polygon boundaries. The pointer to the boundary
index is written only in the second pass.

To accurately identify all boundary pixels, we use the \textit{conservative
rasterization} feature provided by the graphics API.  Conservative rasterization
identifies and draws all pixels that are touched by a line (or triangle), and is
different from the default rasterization, wherein pixels not satisfying the
necessary intersection condition are not drawn.

\myparagraph{Optimizing for Rectangular Range Queries.}
The constraints for these queries consist of axis parallel rectangles.
Thus, given a rectangle, it is sufficient to store its diagonal endpoints.  To
create a canvas, we first convert each rectangle into two triangles using the
geometry shader, and then use the same fragment shader that was used for polygon
primitives above.

\myparagraph{Canvases for Distance-Based Queries.}
Distance constraints are converted into polygonal constraints.
In particular, we use geometry shaders to create these (query) polygons as follows.
When the query is with respect to the \emph{distance $r$ from a point},
the constraint is a circle which is generated by: 1) 
using the point coordinate and distance $r$ in the geometry shader to
create a square of size $2r$ centered at the point, and then 2) the fragment
shader identifies and draws the interior and boundary of the
required circle  within this square.

\begin{figure}[t]
	\centering
	\includegraphics[width=0.6\linewidth]{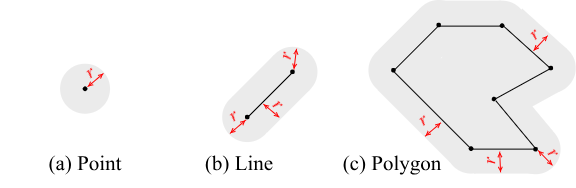}
	\caption{Distance canvases for different primitive types.}
	\label{fig:distance-canvas}
\end{figure}

In the case of \emph{lines}, a ``rounded rectangle" is generated using the
geometry shader. Given a line and a distance $r$, this polygon corresponds to
a rectangle parallel to the line together with semicircles at the two endpoints
as illustrated in \fig{fig:distance-canvas}(b).
For \emph{polygons}, the query constraint is generated by
drawing the polygon interior followed by the boundary lines using the
same geometry shader used above for lines (see \fig{fig:distance-canvas}(c)).
This construction ensures that all points that fall inside the generated polygon
will have their closest distance less than or equal to the distance constraint.

Computing distances to such complex geometries (such as in
\fig{fig:distance-canvas}(c)) is traditionally a costly operation, due
to which existing systems (\eg GeoSpark~\cite{geospark}) typically define the
distance to a line (or polygon) based on the distance to its center.
On the other hand, \textit{\spade can accurately execute these queries}---the
only overhead is canvas creation which is small due to the use of
geometry shaders.

\subsection{Canvas-Specific Indexes}
\label{sec:canvas-index}

We propose two index structures that can be used together with
the canvas-based spatial model to speedup query processing.
In Section~\ref{sec:query-engine}, we also show how the GPU-friendly
operators defined in \cite{gpu-algebra} can be used to create
these indexes.

\myparagraph{Boundary Index.}
A common operation across spatial queries is to test whether a given location
intersects a geometric object.
This is trivial if the corresponding pixel on the canvas is completely \textit{inside}
or \textit{outside} the object.
However, if the location falls on a boundary pixel,
it is necessary to explicitly compute the intersection to enable accuracy.
This is typically accomplished using costly point-in-polygon, 
line-polygon, or polygon-polygon intersection tests, each of which
have time complexity polynomial in the size of the primitives (i.e., the number of vertices making up the primitive).

The first index we propose is the \textit{boundary index} which serves
two important purposes: 
1)~enable boundary tests necessary for accurate queries; and 
2)~speed up spatial intersection tests.
The index is defined as a lookup table where the entries 
correspond to a geometric primitive that can be used to verify whether a
point falling within a boundary pixel of the canvas intersects the geometry.

For points and lines, this lookup table is trivially defined: the point
coordinates and the line end-point coordinates form the necessary entries, and
the $v_b$ value of the boundary pixel of a canvas points to the corresponding
entry in the table.
In other words the data itself becomes the boundary index.

In the case of polygons, the lookup table stores the triangles making up the polygon 
together with the unique identifier of the geometry.
In addition, we also store a mapping that associates every edge of a polygon to the
triangle incident on it.
For example, consider the polygon $P$ in \fig{fig:selection-query}. 
It consists of 4 boundary edges, of which $e_1$ and $e_2$ map to the triangle $t_1$,
while the other two edges map to triangle $t_2$.
This mapping is used during canvas creation to associate each boundary pixel (part of a polygon edge)
to the corresponding triangle.
Now, if a geometric primitive intersects with a boundary pixel,
to test whether it also intersects with the corresponding polygon,
it is sufficient to simply test if that primitive intersects the triangle indexed 
by that boundary pixel.
For example, to test whether the cyan point intersects $P$ in \fig{fig:selection-query},
it suffices to simply test whether this point intersects $t_1$.
Thus, costly point-in-polygon, line-polygon, and polygon-polygon intersection tests 
now become \textit{constant time} point-triangle, line-triangle, and triangle-triangle tests, respectively.
Brinkhoff~et~al.~\cite{Brinkhoff94} had tried out similar index structures where the polygons were
partitioned into either smaller convex polygons, trapezoids, or triangles, which were then
stored using the R-tree. As they mention, this would still require $O(n)$ time in the worst case.

\myparagraph{Layer Index.}
Recall that each geometric object is represented using one canvas. Thus, when working with data
involving millions of objects, the operators need to be executed on millions
of canvases.
The goal of the 
\textit{layer index} is to reduce the number of canvases 
created and used, and thus improve the query performance.
In addition, it also improves the GPU occupancy since it allows multiple objects to be
processed simultaneously.
It is primarily defined for line and polygon primitives. 
Given a set of geometric objects (lines or polygons) that form a canvas, this index
partitions the data into a set of layers, such that no two geometric objects in a given layer intersect.
Note that in the worst case, when all objects intersect, each object is in its own layer which 
results in the same performance as having no layer index.
However, such a scenario is rare in real world data, thus providing significant
query speedup in practice.

%% file: operators.tex
\section{Spatial Query Engine}
\label{sec:query-engine}

The spatial query engine handles planning, optimizing, and executing the
queries. In this section, we first discuss the shader-based implementation of
the algebraic operators that are used for different queries--for
their formal definitions, see \cite{gpu-algebra}.
We then describe in detail the query evaluation and optimization strategy 
followed for the different spatial queries.
Finally, we show how the canvas-specific indexes introduced above
can be computed using the GPU operators.

\subsection{Implementing the Algebra Operators}
\label{sec:gpu-operators}

\begin{figure}[!ht]
\centering
\includegraphics[width=0.75\linewidth]{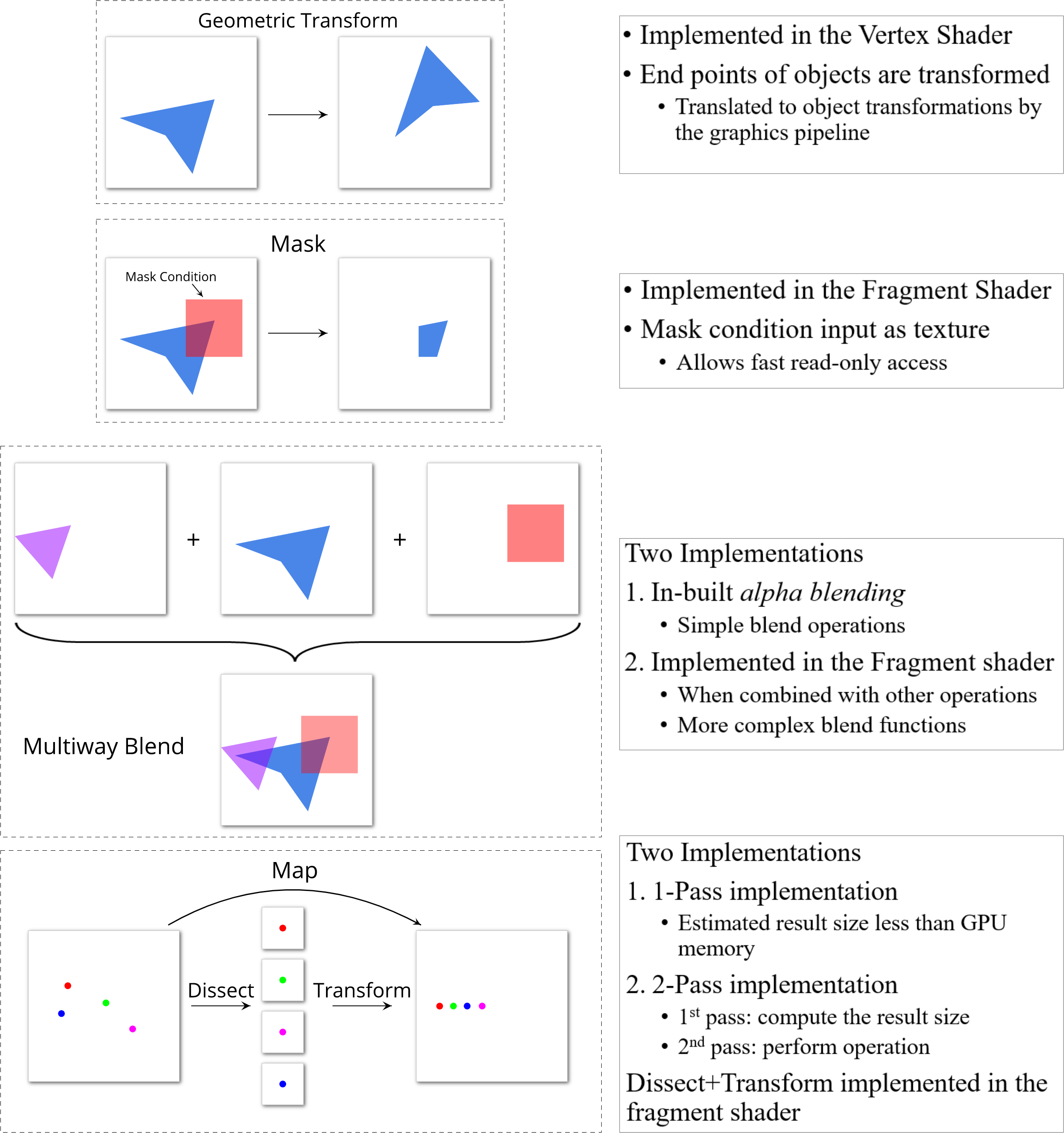}
\caption{Operators implemented in \spade}
\label{fig:operators-summary}
\end{figure}

\spade implements the following four operators that are summarized in \fig{fig:operators-summary}.

\myparagraph{Geometric Transform.}
When a canvas is created, the vertex shader performs the geometric
transformation on the coordinates of the geometries to 
convert them into screen space. Here, the screen space is defined
based on the query inputs and parameters (\eg bounds of the selection query polygon),
thus clipping (and not processing) regions that are not touched by the query (\eg see \fig{fig:selection-query}).
It is also used to
change the coordinate system of the underlying data, in particular, from 
the latitude-longitude based coordinates to the meters-based EPSG:3857
coordinates for distance and \knn queries.

\myparagraph{Mask.}
The mask operation is implemented in the fragment shader.  Fragments are
processed in parallel by the fragment shader, where they are tested against the
mask condition. When this condition corresponds to a polygonal constraint,
the corresponding constraint canvas together with the boundary index is used by the shader 
to test the mask condition  (\eg see \fig{fig:selection-query}).  If the fragment 
  falls on a boundary pixel, then the appropriate boundary tests are
  performed. 
  To reduce memory lookup latency, \spade stores the constraint
  canvas in the part of the GPU memory that is segregated for textures.  Since
  textures are read-only, hardware optimizations allow faster data access as
  compared to traditional global memory.

\myparagraph{Multiway Blend.}
Since spatial \datasets are made up of multiple canvases, a single multiway
blend operator becomes beneficial compared to executing multiple blend
operations. Hence, \spade implements the multiway blend operation.
For simple blend functions such as additions (used for aggregations),
we use the API-provided \emph{alpha blending} operation.
When the blend function is more complex, or in cases where blend is combined with other operations,
we perform the blending in the fragment shader.
The latter scenario is common when a blend is followed by a mask in a query plan. 
Since the operations are executed by explicitly rendering (drawing) the appropriate data,
%
instead of having two (or more) rendering passes, one to perform blend and one to perform
mask, combining the two operations into a single fragment shader helps avoid unnecessary
rendering passes. 

\myparagraph{Map (Dissect + Geometric Transform).}
Since the \textit{dissect} operation is never used in isolation in any of the queries, 
  we chose to implement the \textit{Map} operation as a single operator that
  performs a dissect followed by a geometric transform.
By definition, the \textit{Map}  operation creates a canvas with a single point for each non-null 
fragment of a canvas, where the coordinates of the point are determined by the
fragment values. This results in the creation of numerous canvases, and a na\"{i}ve implementation
becomes expensive.

One approach to overcome this is to \emph{implicitly} represent each generated canvas within a 
single output canvas. Since each of the generated canvases has only one valid point, 
if we can store each of these points using unique pixels, then a single output canvas is sufficient.
We accomplish this by using the identifiers associated with the generated canvases 
to uniquely encode the pixels in the output. 

We have two distinct implementations of the Map operation, and the query optimization strategy
chooses the appropriate implementation at runtime based on the query parameters:

\myparagraphem{1.}~Given estimate $n_{max}$, the
maximum possible count of points that are generated by the dissect operation
(described in the following section), 
a canvas with texture size (resolution) equal to $n_{max}$ is created,
and the Map operation stores each created point at a unique location within this canvas.
This canvas can be thought of as a list of size $n_{max}$ having both null and non-null values. 
When the query is complete, a GPU-based parallel scan operation~\cite{harris2007parallel} is 
executed (to remove the null elements of the list) and return the result.

\myparagraphem{2.}~When the maximum count cannot be estimated,
then the Map operation is performed in two steps.
The first step performs a simulated Map operation, counting the number of points
created. This count is then used in the second step to perform the actual Map operation.
Note that the second step may require multiple iterations depending on the count of the points
computed in the first step.

\subsection{Query Evaluation}
\label{sec:query-execution}

\spade closely follows the query plans described in \cite{gpu-algebra, gpu-algebra-full} for the different
spatial queries. 
Next, we describe
the design choices made in our implementation
for the different query types when the data fits into 
GPU memory.
The following section details how this is extended
for off-core processing of large \datasets.

\begin{figure}[!t]
	\centering
	\includegraphics[width=0.6\linewidth]{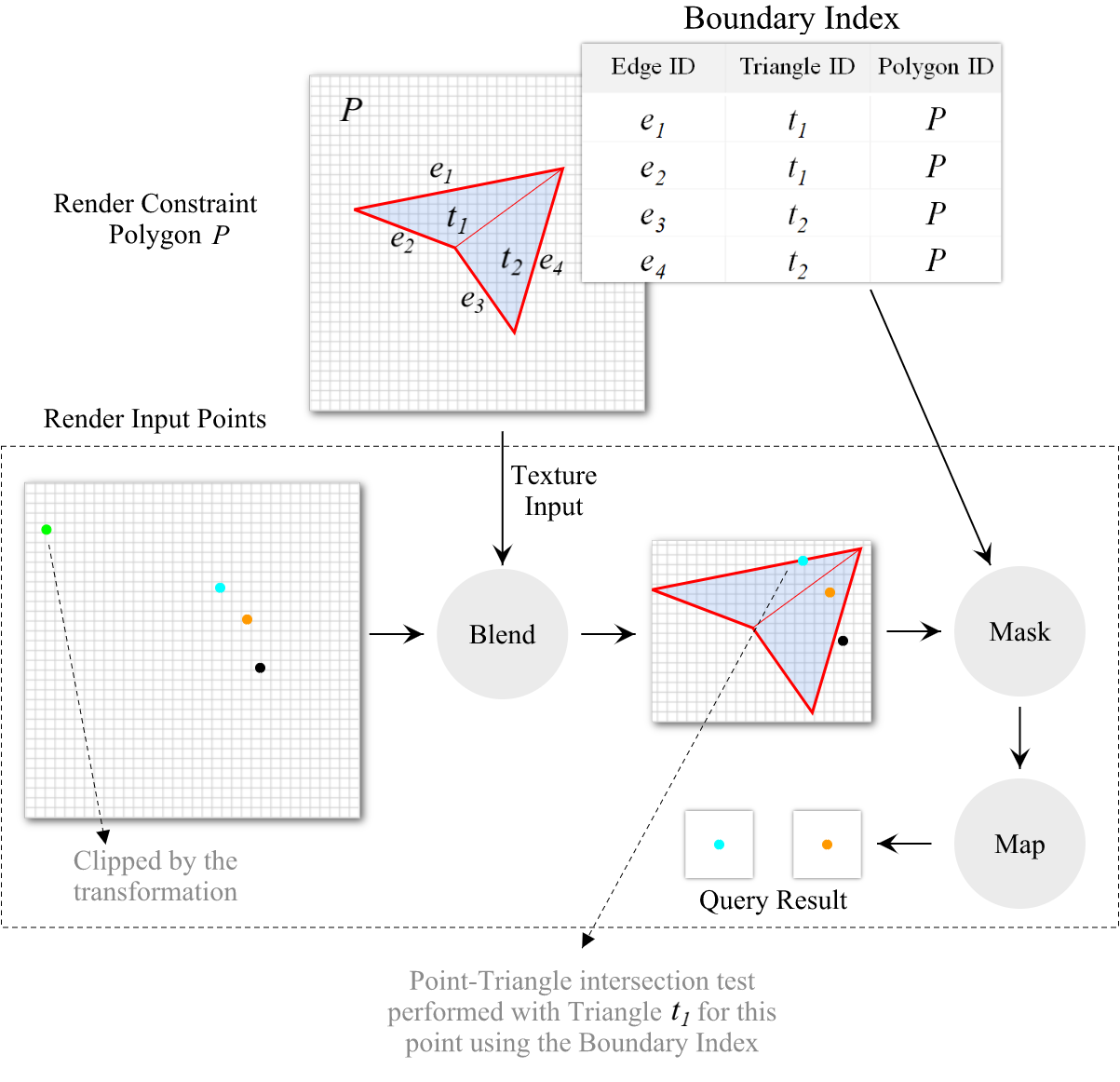}
	\caption{The steps executed for a selection query. 
		The triangulated polygon is rendered and passed as a texture during the point rendering phase.
		As each point is rendered, if it falls outside the bounding box of the
		polygon (green point), it is clipped.
		Points that are not clipped  then undergo the blend, mask, and map operations in the fragment shader.
		During the mask operation, if the point (cyan point) falls on a boundary pixel (bold red lines), then a point-triangle intersection
		test is performed using the boundary index.}
	\label{fig:selection-query}
	\vspace{-0.1in}
\end{figure}

\myparagraph{Spatial Selection.}
A spatial selection query identifies all geometric objects from a given query \dataset
that satisfies a spatial constraint and a select condition. Here the geometric objects can be points,
lines, or polygons, or any combinations of these.
\spade currently supports \textit{intersection} (analogous to the SQL ST\_INTERSECTS condition),
when the spatial constraint is a polygon.\footnote{A rectangular range is also treated as a polygon.}
In Section~\ref{sec:discussions}, we  discuss the use of the
containment (SQL  ST\_CONTAINS) condition.
The original query plan used for spatial selection~\cite{gpu-algebra} performs 
a blend followed by a mask operation to generate a set of result canvases.
To extract the query result from the result canvases, 
we use a modified plan
that performs a Map operation following the mask. 
This query is implemented in three steps as illustrated in \fig{fig:selection-query}:
\vspace{-0.1in}
\begin{enumerate}[leftmargin=12pt] \denselist
	\item Create a canvas for the polygonal constraint and store it as a texture. This is done in one rendering pass.
	
	\item Create canvases for the query \dataset. 
	This is also accomplished in a single rendering pass as follows:
	All geometric objects are simultaneously drawn in this rendering pass. 
	Each fragment\footnote{Even when multiple objects overlap, overlapping fragments are still processed by shaders.
		Thus, there is no loss of data even in such scenarios.}
	 is subjected to the blend and mask operation, and if the mask condition is satisfied, then the Map operation 
	 stores the data onto an output canvas. This way, the object canvases need not be stored and can be discarded
	 immediately after the fragments are processed.
	 The query constraint canvas, stored as a texture,
	 is used by the fragment shader to accomplish the mask operation.
	 Only objects within the query region defined by the 
	 bounding box of the constraint polygon are processed in this step.
	 This is accomplished by defining a transformation of the input that clips (discards) any object outside this extent,
	 and thus avoiding unnecessary processing. 
	 
	 \item The scan algorithm is used to extract the query result from the output canvas.
\end{enumerate}

\myparagraph{Spatial Joins.}
A spatial join between two spatial \datasets, $D_1 \bowtie D_2$ is implemented as a collection of spatial selections,
where the selection constraint is from one of the \datasets. 
The in-memory based implementation of the join query uses the \textit{layer index}, introduced earlier.
We look at how this is accomplished for the following two join scenarios:

\myparagraphem{(1) Polygon $\bowtie$ Point:} Let $D_1$ be the polygon
	 \dataset and 
	$D_2$ be the point \dataset. Due to the layer index, each layer of $D_1$ contains polygons that do not intersect, 
	and thus can be  represented in a single canvas texture. Let there be $l$ layers 
	in the layer index of $D_1$. The implementation then performs $l$ select operations 
	one for each layer. 

\myparagraphem{(2) Polygon $\bowtie$ Polygon:} Here, both $D_1$ and $D_2$ are polygonal \datasets.
	Let $l_1$ and $l_2$ be the number of layers respectively in the two \datasets. 
	Without loss of generality, let $l_1 \leq l_2$. Similar to the previous scenario, 
	$l_1$ selection operations are executed for each of the layers in $D_1$. 

\myparagraph{Distance-Based Queries.}
Distance-based selections and joins differ from spatial selections and joins only in 
Step~1: the constraint canvases are created using the geometry shader as defined in Section~\ref{sec:canvas-creation}.

We support two types of distance-based joins, which are also executed similar to spatial joins:

\myparagraphem{(1)}~Given \datasets, $D_1,D_2$, and a distance $r$, the join returns all pairs of objects $(x_i, y_j)$
	such that $x_i \in D_1$, $y_j \in D_2$ and $\text{distance}(x_i,y_j) \leq r$. The \dataset with the smaller number of 
	elements is used for creating the constraint canvases.
	
\myparagraphem{(2)}~Given \datasets, $D_1,D_2$, and a set of distances $\{r_1, r_2, \ldots, r_n\}$, $n = |D_1|$, the
	join returns all pairs of objects $(x_i, y_j)$ such that $x_i \in D_1$, $y_j \in D_2$
	and $\text{distance}(x_i,y_j) \leq r_i$. In this case, constraint canvases are created on $D_1$.

Since the distance parameter of the join query is provided together with the query, unlike polygonal \datasets, 
it is not possible to have the layer index built beforehand -- instead the index
is built on the fly during query execution. 

\myparagraph{Spatial Aggregations.}
We support two implementations for the aggregation query.
The first performs the necessary join (or select) followed by the count operation which is accomplished using 
a geometric transformation followed by a multiway blend. Here, the results from the select (or join) are moved
to a unique location by the geometric transformation, and the blend operation counts the objects in each of these
unique locations to obtain the aggregation result.
The second implementation, follows the alternate query plan described in
\cite[Section~5.2]{gpu-algebra-full}. This plan is customized for aggregations over point data, 
and uses the multiway blend, mask, and map operators. This plan first computes
partial aggregates of the data using multiway blend, which are then 
combined together using the mask and map operations to generate the final result.
Query speedup is obtained in this scenario by avoiding the materialization of the join (or select) results.
This is possible only for points since any given point can be in at most one pixel of the canvas.
Note that the latter implementation is always chosen by the optimizer when working with point \datasets.

\myparagraph{\knn Queries.}
Since common real-world \knn queries as well as existing spatial engines support \knn queries only over point data,
we chose to customize the query plan used for \knn queries with respect to point data as well.
Given a spatial \dataset $D$, consider a \knn selection query that identifies $k$ points closest to
a query point $p$. 
Let $B$ be the bounding box of $D$. Let $r_{max}$ be the maximum distance from $p$ to the endpoints of $B$.
The \knn query is then executed in 3 steps:
\begin{enumerate}[leftmargin=12pt] \denselist
	\item A circle \dataset $C$ with $c$ circular constraints is generated, where every circle has center $p$,
	and $i^{\text{th}}$ circle has radius $r_i = \frac{r_{max}}{\alpha^{i}}$, $0 \leq i \leq c$ and $\alpha$ is a constant $> 1$. 
	Distance-based canvases are used for this purpose. 
	A spatial aggregation query is then run on $D$ using $C$ as the constraints.
	Note that such an approach might seem inefficient from a traditional CPU algorithmic perspective.
	However from a computer graphics perspective, it is designed to harness the GPU as follows.
	The plan used for aggregation query above requires just one pass over the polygon constraints 
	irrespective of the number of constraints in the query. In terms of graphics operations, 
	this amounts to drawing the set of polygons once. GPUs are primarily designed for this 
	operation\footnote{GPUs can render million polygons at rates as high as 60fps}. 
	Since 
	(i)~polygons in this case are simple circles rendered efficiently using geometry shaders (Section~\ref{sec:canvas-creation}); and 
	(ii)~the number of circles created is logarithmic on the spatial extent of the input in terms of the canvas resolution, 
	which in practice is at most a few hundreds, the performance overhead of this step is small.
	\item Let $r_i$ be the radius of the circle such that aggregation$(c_i) \geq k$ and aggregation$(c_{i-1}) < k$.
	A distance-based selection is now run on $D$ with $r_i$ as radius. 
	\item The query results are then sorted based on the 
	distance to point $p$, and the $k$ closest points are selected from the sorted list.
\end{enumerate}

A \knn join is executed similar to the \knn selection query. 
Let $D_1$ and $D_2$ be the input \datasets. Let the query require identifying the $k$ nearest neighbors of points in $D_1$.
Then, the circle \dataset containing circles with varying radii is created for each of the points in $D_1$ in the
first step to identify the appropriate radius for each point in $D_1$.
Then the Type~2 distance join is performed using the computed radii in Step~2. Finally, similar to the \knn selection query, 
the query results are sorted based on distance and the required result is computed.

\subsection{Out-of-Core Queries}
\label{sec:ooc-queries}

We now discuss how the above query strategies are extended to data that do not fit in
GPU memory (and may or may not fit in CPU memory).

\myparagraph{Grid-based index filtering.}
\spade uses a clustered grid-based indexing structure where each grid cell
corresponds to a block of data that falls into that cell. The block size
is tuned based on the system configuration to ensure that
any given grid cell fits into GPU memory 
(see Section~\ref{sec:setup}).
The collection of non-empty grid cells is stored as a set of bounding polygons, 
\ie for each cell, instead of storing simply the bounding box, we store the
convex hull over the geometries present in that cell.
Thus, the index itself forms a polygonal \dataset.
When an object spans more than one cell, it is added to the cell that contains its centroid 
and the cell’s boundary is expanded. Thus, the cells themselves could intersect unlike in a traditional grid index. 
The filtering is then accomplished by executing the appropriate selection/join query on this set of boundary polygons.
Thus, having intersecting cells does not affect the query evaluation strategy.

Due to the use of more detailed polygons instead of bounding rectangles, \textit{the index filtering stage of query execution
filters out more data, thus reducing the amount of data used for the refinement stage}.
Such a strategy is not used in a traditional setting due to the high processing cost this filtering incurs.
However, the overhead is low in our case since this representation enables 
the reuse of GPU-based selections and joins to perform fast index filtering.

Next, we describe the query execution strategy for spatial selections and joins. 
Other queries are also executed using a similar strategy. 
Note that when the data does not fit in CPU memory, the cells of the index 
are memory mapped and loaded into the CPU as
and when necessary

\myparagraph{Spatial Selection.}
In the index filtering phase of the query, a spatial selection query is first
executed on the bounding polygons (corresponding to the cells) of the clustered grid index
to identify all grid cells that satisfy the selection constraint.
For the refinement phase, the GPU in-memory spatial selection query is then executed on each of these
grid cells.

\myparagraph{Spatial Joins.}
\spade incorporates two strategies to perform spatial joins over out-of-core data.
The first makes use of the layer-index based in-memory join described above as follows.
Let $D_1$ and $D_2$ be the two \datasets used for the spatial join.
The filter phase first performs a Polygon $\bowtie$ Polygon
join over the bounding polygons of the indexes corresponding to
$D_1$ and $D_2$ respectively.
This join returns all pairs of grid cells $(g_i^1,g_j^2)$, where
$g_i^1 \in D_1$ and $g_j^2 \in D_2$.

A spatial join is then executed for each pair of grid cells.

The second strategy follows a na\"ive approach that implements it as a loop of selects. 
Even though the use of the layer index decreases the number of canvases involved in the join,
since multiple polygons are part of a single layer (and hence canvas), these layers can cover a large region.
Say, \dataset $D_1$ contains such large-extent layers, then for each layer,
data corresponding to large extents from $D_2$ needs to be transferred to the GPU.
On the other hand, if the individual polygons in $D_1$ cover a relatively smaller region, then
depending on the distribution of the data in $D_1$ and $D_2$, the total data transferred to the GPU can be significantly 
lower when summed over all polygons in $D_1$. 
While the layer index-based approach optimizes for the number of rendering passes, 
it does not consider the memory transfer. Since the data transfer forms the primary bottleneck in
query execution times, the na\"ive strategy can therefore perform better in such scenarios.

\subsection{Query Optimization Strategies}
\label{sec:query-optimization}

\spade implements a 
query optimizer~(QO) that performs the following tasks:

\myparagraph{Select the appropriate Map operator implementation.}
The Map operation has two implementations as described above. It is primarily used to 
consolidate the results of a query. When the estimate of the query result ($n_{max}$) is small enough 
to fit into the maximum memory allocated for a single canvas, then the 1-pass Map implementation
is used. Otherwise, the 2-pass implementation is used.
The result estimation for the different queries is performed as follows:

\myparagraphem{1. Selection query:}
Since each object can either satisfy the query constraint or not, the maximum 
size of the output, $n_{max}$, is equal to the number of objects.

\myparagraphem{2. Join query:}
First, consider Polygon ($D_1$) $\bowtie$ Point ($D_2$) joins. Let the number of polygons in 
$i^{\textrm{th}}$ layer of the $D_1$'s layer index be $m$, and number of points in $D_2$
be $n$. The estimate for $n_{max}$ in this case is $n$ 
since no two polygons intersect within a given layer, and a point can intersect only one polygon.

In case of Polygon ($D_1$) $\bowtie$ Polygon ($D_2$) joins,
let the number of polygons in $i^{\textrm{th}}$ layer of $D_1$ be $m$, and number of polygons in $D_2$
be $n$. Unlike the previous scenario, the estimate of $n_{max}$
is now $n \times m$ (a single polygon in $D_1$ can intersect multiple polygons in $D_2$).

\myparagraph{Choose the join implementation.}
Recall that when a join query is invoked on \datasets that do not fit in memory,
\spade supports two join implementations: a na\"ive approach of executing a set 
of selections, and using the layer index.
To choose the implementation to be executed, the
QO first computes 
the amount of data that
needs to be transferred to the GPU.
For the na\"ive approach, this is equal to summing the physical size
of the grid cells returned by running the filtering step corresponding
to each of the polygons (\ie a join between polygons in $D_1$ and
bounding polygons of the grid index of $D_2$).
When the layer index is used, the required memory is the sum of the sizes of the grid cells returned 
from the filter step taking into account the loop order of the join (see next).
The join strategy that requires the least memory transfer is then selected
and the corresponding refinement phase is then executed.

\myparagraph{Identify the order of join operations.}
Given a join query, the QO executes this query as a collection of
$l$ selects, where each select constraint corresponds to a layer 
of the \dataset $D_1$ (or polygons if the na\"ive approach is used). 
Based on the index filtering using the selection constraint, the appropriate 
grid cells of $D_2$ need to be loaded into the GPU for each of the selects.
The main idea is for the QO to use this data transfer overhead 
to decide on the order of join operations. 
We chose this simple measure because,
as we show later in the evaluation, the data transfer
time forms the primary bottleneck during query execution.
In the current implementation, the QO orders the operations such at least
one grid cell or layer (in either $D_1$ or $D_2$) is common 
between consecutive selects, thus trying to share the memory 
transfer from one iteration of the join loop to the next.

\vspace{-0.15in}
\subsection{Canvas-Specific Index Creation}
\label{sec:gpu-index}

\vspace{-0.15in}
\myparagraph{Boundary Index.}
The boundary index has to be created only for polygonal \datasets.
Let $P_b$ be the set of lines that make up the boundary of the polygon data.
Let $P_t$ be the set of triangles obtained through polygon triangulation.
The required boundary index is then computed by 
executing a spatial join using $P_b$ and $P_t$ 
as the data sets. Note that in this case, each triangle 
is its own polygon, and hence this data forms a trivial boundary index
for the purpose of this join.

\myparagraph{Layer Index.}
Creating a layer index is accomplished through an
iterative process that makes use of the GPU operators.
Consider the $i^{\text{th}}$ iteration. When this iteration is executed, $i-1$ layers are already computed.
Let $\C_{rem} = \{C_1, C_2, \ldots C_n\}$ be set of canvases that have not yet been assigned a layer after $i-1$ iterations.
Then the polygons in the $i^{\text{th}}$ layer are computed in two passes as follows:

\myparagraphem{Pass~1:}
A multiway blend operation is applied on $\C_{rem}$, where the blend function 
between two objects is defined such that when two objects overlap, it
removes the overlapping regions of the object having a lower identifier.
Let $C_{max}$ be the canvas resulting from the above multiway blend operation.

\myparagraphem{Pass~2:}
This pass first performs a blend between $\C_{rem}$ and $C_{max}$ followed by a mask
to identify and discard objects that have been cropped in Pass~1.
The objects remaining after the mask operation is used for the $(i+1)^{\text{th}}$ iteration, and the
ones that were discarded are added to the $i^{\text{th}}$ layer.

%% file: exp.tex
\begin{table*}[t]
	\centering
	\ssmall
	\begin{tabular}{|c||l|l|l|l|l|l|}\hline
		&  & \multicolumn{1}{c|}{\textbf{Spatial}} & \multicolumn{1}{c|}{\textbf{\#}} & \multicolumn{1}{c|}{\textbf{\#}} & \multicolumn{1}{c|}{\textbf{Physical}} & \\
		\textbf{Name}  & \multicolumn{1}{c|}{\textbf{Type}} & \multicolumn{1}{c|}{\textbf{Extent}} & \multicolumn{1}{c|}{\textbf{Objects}} & \multicolumn{1}{c|}{\textbf{Points}} & \multicolumn{1}{c|}{\textbf{Size}} & \multicolumn{1}{c|}{\textbf{Description}} \\ \hline \hline
		Taxi Data~\cite{nyctaxi} & Points & NYC & 1.22~B & 1.22~B & 29~GB & Pickup coordinates from taxi trips that happened from 2009-2015\\ 
		Tweets & Points & USA & 2.28~B & 2.28~B & 63~GB & Geo-locations of tweets gathered using the Twitter API.\\ \hline
		Neighborhoods~\cite{nycneighs} & Polygons & NYC & 195 & 105.0~K & 2.7~MB & Neighborhood boundaries of NYC. \\ 
		Census~\cite{nyccensus} & Polygons & NYC & 2,165 & 156.7~K & 4.0~MB & Census tract boundaries of NYC. \\ 
		Counties~\cite{uscounties} & Polygons & USA & 3,109 & 5.33~M & 134~MB & Boundaries of counties within the USA.\\ 
		Zip Code~\cite{uszip} & Polygons & USA & 32,657 & 50.53~M & 1.3~GB & Boundaries of all zip code regions within the USA. \\ 
		Buildings & Polygons & World & 114~M & 764~M & 19~GB & Building outline polygons from Open Street Map that was used in \cite{pandey@vldb2018}. \\ 
		Countries~\cite{world} & Polygons & World & 250 & 215.5~K & 5.4~MB & Boundaries of all countries. \\ \hline
	\end{tabular}
\caption{\Datasets used in our evaluation. A CSV file with only the coordinates was used to compute the size of the point \datasets, while the files in WKT format was used for polygonal \datasets (some of these \datasets consists of multi-polygons).}
\label{tab:datasets}
\end{table*}

\section{Experimental Evaluation}
\label{sec:exp}

In this section, we first perform an experimental evaluation to assess the efficiency and effectiveness
of \spade using real-world \datasets. 
We then assess the scalability of \spade across query sizes and data distributions using synthetically generated data.
Our main goals are: (1)~test the
feasibility of using a laptop for spatial queries over large \datasets; and
(2)~assess the scalability of \spade for different workloads.

\subsection{Experimental Setup}
\label{sec:setup}

\myparagraph{\spadebf Implementation.}  \spade was implemented using
C++ and OpenGL\cite{redbook}, thus allowing it to work on any GPU and
on different operating systems.  We chose OpenGL over the more recent
Vulkan~\cite{vulkan} since Vulkan drivers are still nascent for many
GPUs.  Since the OpenGL shaders also works on Vulkan, it is easy to port \spade
to Vulkan in the future.

\myparagraph{\DataSets and Queries.}
Table~\ref{tab:datasets} summarizes the \datasets used in the
evaluation.  To assess the performance of the system,
we chose \datasets of varying sizes and 
spatial extents.
In addition, to test SPADE's ability to handle large data, we include \datasets  that are
larger than the main memory of commondity machines. 
Note that some of these \datasets (e.g., taxi, tweets) are larger than
those used in a recent evaluation of cluster-based
systems\cite{pandey@vldb2018}. 

The workload includes queries with different selectivity (controlled
by the regions defined for selection and join) and reflects
queries used in  real applications. 
While we only use point and polygonal \datasets in our experiments,
note that the performance of queries over 
polygonal \datasets can be used as a worst case upper bound
for (poly)line \datasets---drawing lines and performing line-intersection tests 
is cheaper than drawing polygons and performing triangle-intersection tests.

\myparagraph{Approaches to Spatial Query Evaluation.}
Approaches for spatial data analysis can be classified into three groups
based on their target environment:
large main-memory servers, map-reduce clusters, and laptops/desktops.
We compare \spade against representatives of these groups
that were shown to be efficient in recent experimental studies.
Note that the query times across these approaches are incommensurable---it 
is not our intent to perform a head-to-head comparison between
\spade and these systems. Our goal is to obtain points of reference to
assess the suitability of \spade as an alternative to fill the
existing gap for spatial analytics on commodity hardware as
well as better understand the trade-offs involved.

\myparagraphem{1. Main-memory servers:} Common spatial libraries are
limited to in-memory processing
and thus require large main-memory servers to handle big
\datasets.
We selected the Google S2~geometry library~\cite{google-s2} as a
baseline for the performance on large-memory servers, since a recent
study reported that it has the most consistent performance across query classes~\cite{pandey@dsc2020}.
Furthermore, it is optimized for spherical geometry, and is thus suitable to
handle the \datasets used in this evaluation.

\myparagraphem{2. RDBMS:} We experimented with PostgreSQL and a
commercial database system, and found that these systems performed
significantly slower than the in-memory S2-library. For example, a
selection query on the taxi data that took 5m 21s using the commercial
system took only 50s using S2.  Similarly, a taxi-neighborhoods join
took close to 3 hours on the commercial system, but only 7m 33s
using S2.  Moreover, PostgreSQL performed slower than the commercial
system.  Therefore, we omit a comparison with these systems.

\myparagraphem{3. Specialized GPU-based approaches:} Several approaches
have been proposed that harness the power of GPUs to support specific
queries and/or applications (see Sections~\ref{sec:intro} and
\ref{sec:related-work}).
Since these are highly optimized, they can provide an upper bound for
the performance of the query class they support. As a representative
of this class, we selected 
the open-source GPU-based STIG~\cite{STIG2016} library, which allows the execution of
spatial selection queries over large point \datasets even on laptops.
For our evaluation, we used
the standalone version that can be executed independently without
MongoDB~\cite{stig-standalone}.

\myparagraphem{4. Cluster-based systems:} These systems are
specifically designed to handle large data sizes.
As a representative of this class, we selected
GeoSpark~\cite{geospark}(v1.3.1) for two main reasons: in a recent
experimental evaluation of spatial analytics systems,
Pandey et al.\cite{pandey@vldb2018} found that
GeoSpark~\cite{geospark} outperformed the other cluster-based systems;
and it supports a rich set of queries.
We tuned GeoSpark for every query that was run (see below). 
Since having more nodes in the cluster can increase memory transfer overheads, 
to be fair, we do not include the setup (\eg the time needed to partition and copy data to
cluster nodes) and indexing time--we report only the execution time 
after the data is already loaded and indexed in the cluster nodes.
This setup time is around half a minute for the smaller polygonal \datasets,
and it takes 2 and 4 minutes for the larger Taxi and Twitter data respectively.
However, note that there could still be an overhead when the query results
from the different nodes are combined.

\myparagraph{Configuration.}
\spade and STIG were run on a \emph{laptop} having an Intel Core i7-8750H
processor, 16~GB memory, 512~GB SSD, and a 2~TB external SSD 
connected via a Thunderbolt~3 interface. 
The laptop is equipped with a Nvidia GTX~1070 Max-Q GPU with 8~GB graphics memory.
S2 was run on machine with a 2.40 GHz Xeon E5-2695 processor and 256 GB of RAM.
GeoSpark was executed on a cluster with 17 compute nodes, each node having 256~GB
of RAM and 4~AMD~Opteron(TM)~6276 processors running at 2.3GHz. 

\myparagraph{Database Setup and Tuning.}
To evaluate the performance of \textit{STIG}, indexes were built on the Taxi
and Twitter \datasets. The \textit{leaf block size} was set to 4096 since it 
gave the best performance compared to other sizes.

We used the S2PointIndex and S2ShapeIndex for point and polygon data respectively,
when using the S2 library.

The performance of \textit{GeoSpark} depends on multiple parameters: number of 
partitions (of the SpatialRDD), spatial partitioning strategy, the spatial index used,
as well as the \dataset order (in case of join queries).
We executed each query multiple times, trying all possible combinations, and
used the setting that resulted in the fastest execution time. Unlike the number
of SpatialRDDs, the possibilities for the partitioning strategy, index type, and
\dataset order are limited. Therefore, to choose the number of partitions, we
varied this number from 4 to 128K to identify the best value for a given query.
For all queries, we noticed that the performance degraded after reaching a peak
when the number of RDD partitions became too large.
For point data, the RDD was created using the KDB tree partitioning strategy, while for polygonal 
data, a Quad tree strategy was used. An R-tree index was created on each of the partitions 
for both point and polygon data. 
Similar to \cite{pandey@vldb2018}, we followed the guidelines in \cite{cloudera} 
to tune our cluster. 

The grid cell sizes for creating indexes in \spade were selected based on the
amount of memory present in the GPU. Since all the \datasets used are GIS-based,
we used Open Street Map \textit{zoom levels}~\cite{zoomlevels}
to specify the grid cell sizes. In particular, we restrict the 
zoom level such that the maximum size of the data corresponding to a 
grid cell is less than 2~GB. Note that this includes
not just the coordinates but also the associated metadata and canvas indexes.
This ensures that the GPU can fit two grid cells worth of data, and still has 
around 4~GB free to store intermediate buffers and results. This is sufficient since at any given 
point in the query execution, the GPU processes at most 1 grid cell from every \dataset
associated with a query, and the supported queries can be applied to at most two \datasets (like joins).
For query constraints that are provided during query time (\eg polygonal constraint for selection queries),
the necessary canvas-specific indexes are computed on the fly and index
construction is included as part of the total run time.

\begin{figure*}[t]
	\centering
	\includegraphics[width=\linewidth]{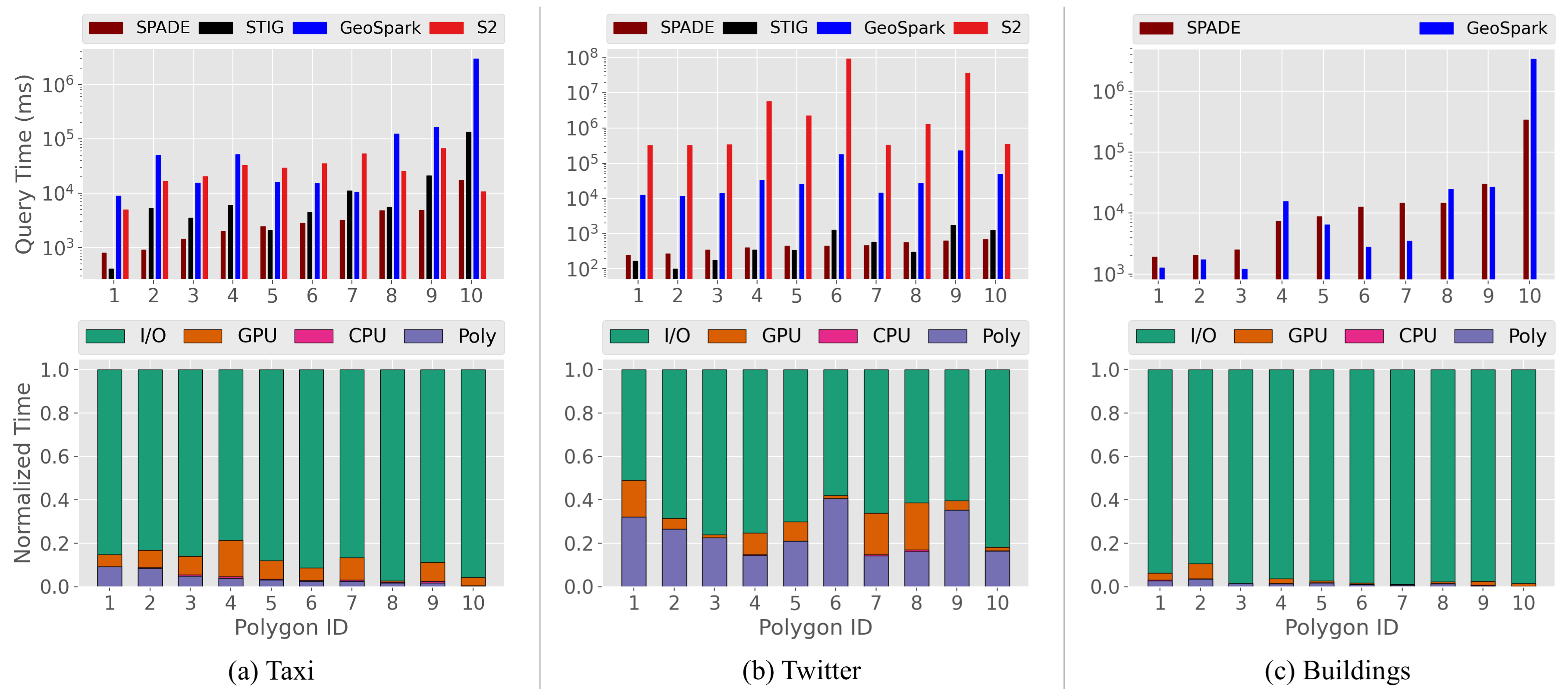}
	\caption{Performance of different approaches to evaluate spatial
		selection queries. The top row shows the total query run time required
		for 10 queries with different polygonal constraints. The polygons are ordered based on the query time using \spade. The query polygons were chosen from (a)~Neighborhoods; (b)~Counties; and (c)~Countries for the three input \datasets respectively.
		The bottom row shows the breakdown of the \spade query
		time. 
	}
	\label{plot:selection}
\end{figure*}

\subsection{Selection Queries}
\label{sec:selection}

We evaluate selection queries using 3 \datasets---Taxi, Twitter, and
Buildings.  Due to space limitation, we discuss the results only for selection
queries with polygonal constraints.

\myparagraph{Polygonal Selection of Points.}
\figs{plot:selection}(a)[top]
and \ref{plot:selection}(b)[top]
 compare
the running times 
for 10 selection queries over the Taxi and Twitter data respectively. 
The polygonal constraints for the Taxi experiments correspond to different 
NYC neighborhood boundaries, while they correspond to U.S. county boundaries for the Twitter experiments.
STIG can be faster than \spade when the running times are less than a few hundred
milliseconds.
This can be attributed to two reasons: 
(1)~\spade incurs an overhead for
triangulating the query polygons and to create the boundary index. It
also incurs an additional overhead in setting up OpenGL states for the
different rendering passes.
When the running times are small, these overheads
are a non-trivial fraction of the processing times.
STIG, on the other hand, has neither a polygon processing overhead, nor an
OpenGL overhead since it uses CUDA; and (2)~the STIG index was designed
primarily to improve the filtering phase of the query, and thus it reduces both the
GPU memory transfer overhead and the number of point-in-polygon tests that need
to be performed. In contrast, \spade uses a much coarser filtering
strategy (recall that each grid cell can store as much as 2~GB worth of
data---in the case of point data, there is no explicit boundary index, and hence
more points can be part of a grid cell). When the query selectivity is small,
the advantages induced by the polygon boundary index is negated by the higher
memory transfer times.

\spade consistently outperforms GeoSpark -- with queries being executed between
2X and 108X faster.
When using GeoSpark, we noticed that the running time was highly dependent on the query selectivity. 
The performance of S2 depends on the input data size--it performs on par with GeoSpark for the Taxi data
and it is significantly slower for the Twitter data. 

\myparagraph{Polygonal Selection of Polygons.}
\fig{plot:selection}(c)[top] compares the performance of \spade and GeoSpark 
for 10 selection queries over the Buildings \dataset. 
Here, the polygonal constraints correspond to the different country boundaries.
In contrast to selections over points, the performance of \spade is on par with
that of GeoSpark.
The better performance of GeoSpark can be attributed to the fact that 
the selectivity of these queries was much smaller than the selectivity
of the point queries. 
At the same time, this \dataset presents a worst case scenario for the 
current indexing design of \spade: to take advantage of 
the boundary index, the polygons need to have a width and height of at least two
pixels. Otherwise, whole polygons would fit within a pixel.
Thus, testing for intersection in such cases will devolve 
to using polygon-polygon intersection tests 
(every triangle incident on the boundary has to be tested).
To avoid this, the zoom levels have to be high for this data resulting
in a very large grid size. This in turn requires processing a large number of 
grid cells, which increases both the number of memory transfers initiated, 
as well as the number of rendering passes, thus increasing the total query time.

Note that: 1)  the memory present on the server was not sufficient to
store the S2 index over the Buildings data, and 2) STIG does not
support polygonal data.
Hence we were not able to perform a comparison
with S2 or STIG.

\myparagraph{Analysis and Discussion.}
An interesting point to note is that the variation seen in the running times of
STIG and GeoSpark are roughly similar. This is primarily because both these
approaches perform point-in-polygon tests over the points resulting from the
index filtering phase, the complexity of which depends on the polygon size. Many
of these tests involve multi-polygons or have polygons containing as many as
hundreds of vertices.  Thus, two queries with the same selectivity need not have
a similar performance---query times also depend on polygon complexity.

We also profiled in detail the time taken by \spade by
breaking down the query execution time into four 
components as shown in \fig{plot:selection}[bottom]:
\begin{enumerate}[leftmargin=12pt] \denselist
\item \io time: The time taken for transferring data from disk to CPU memory,
  and from CPU memory to the GPU. Both these transfers are accomplished through
  memory mapping, due to which it was not possible differentiate between the two
  I/O times. We therefore report the combined I/O times.
\item GPU time: The time spent on the GPU.
\item Polygon processing time: The time spent on triangulating the polygonal constraint and creating the boundary index.
\item CPU time: The time spent on tasks processed by the CPU.
\end{enumerate}
While \io dominates the query time in all three scenarios, this proportion is much more prominent for queries over
the Buildings data (consistently taking over 95\% of the time). 
This is due to the larger number of memory transfers initiated in this scenario.
Polygon processing times form a significant fraction for queries over the
Twitter \dataset.
This happens because
the Counties polygons are larger and more complex (in terms of number of
coordinates making up the polygon)---the average size of the polygonal
constraint used for the Taxi data was 739 points, while the same for the Twitter
data was 5183 points. Even though the average size of the Country polygons was
also large (3984 points), since the total query time was larger in this case compared to the points \datasets, the proportion of the polygon processing time is small.

\subsection{Join Queries}
\label{sec:join}

\myparagraph{Point-Polygon Join.}
Table~\ref{tab:point-poly} shows the running time for join queries between point and polygon \datasets.
\spade performs on par with S2 for join queries, while it is faster than GeoSpark.
Both joins with the Taxi data have a similar result size since the polygonal \datasets cover the same
spatial region. Similarly, both joins with the Twitter data also have a similar result size.
However, it is interesting to note that, given similar result sizes, the running
time of \spade and S2 increases with data size, while that of GeoSpark shows the
opposite trend. For example, even though there are significantly more zip codes
than counties, since they cover the same region (USA), the spatial extent of a
single county is much larger than that of a single zip code. Therefore, the
number of points associated with each county (average selectivity per polygon)
is higher than that of points associated with a single zip code in the join
results. This suggests that the time complexity of GeoSpark is highly
dependent on this average selectivity.

\begin{table}[h]
	\centering
	\footnotesize
	\begin{tabular}{|c|l||l|l|l|} \hline
		\textbf{Join \Datasets} & \multicolumn{1}{c||}{\textbf{Join}} & \multicolumn{3}{c|}{\textbf{Query Time (min)}} \\
		& \multicolumn{1}{c||}{\textbf{Size}} & \multicolumn{1}{c|}{\textbf{SPADE}} & \multicolumn{1}{c|}{\textbf{GeoSpark}} & \multicolumn{1}{c|}{\textbf{S2}} \\ \hline \hline
		Taxi, Neighborhood & 1.22~B & 3.67 & 46.34  & 7.56 \\
		Taxi, Census & 1.22~B & 6.92 & 40.25   &  12.31\\ \hline \hline 
		Twitter, Counties & 2.20~B & 11.37 & 636.2 &  20.26 \\ 
		Twitter, Zip Code & 2.20~B & 47.36 & 68.96  &  32.8 \\ \hline
	\end{tabular}
	\caption{Performance of point-polygon joins.}
	\label{tab:point-poly}
\end{table}

\myparagraph{Polygon-Polygon Join.}
Table~\ref{tab:poly-poly} shows the running time for join queries between
polygonal \datasets.  Ignoring the join (Neighborhood, Census) where the data is
very small, we notice that GeoSpark is in general faster than \spade.  This is
primarily because of the significantly higher \io time (in absolute numbers)
incurred by \spade: (1)~in addition to the polygonal data, \spade also has to
transfer both the boundary index and the layer index; and (2)~the number of grid
cells is higher for polygons (zoom levels used are higher) because we also want
to ensure that most polygons span more than 1 pixel. Thus, different cells might
be processed multiple times, 
resulting in an increase of the
\io time.
The exception here is the join between the Buildings and County \datasets. The reason GeoSpark takes more time in this case is again due to the 
high average selectivity per polygon with respect to the County polygons as discussed above.

\begin{table}[h]
	\centering
	\footnotesize
	\begin{tabular}{|c|l||l|l|} \hline
		\textbf{Join \Datasets} & \multicolumn{1}{c||}{\textbf{Join}} & \multicolumn{2}{c|}{\textbf{Query Time (sec)}} \\
		& \multicolumn{1}{c||}{\textbf{Size}} & \multicolumn{1}{c|}{\textbf{SPADE}} & \multicolumn{1}{c||}{\textbf{GeoSpark}} \\ \hline \hline
		Neighborhood, Census & 4.80~K & 0.065 & 2.9 \\
		Zip Code, Counties & 67.8~K & 98.7 & 17.6  \\ 
		Buildings, Counties & 7.90~M & 179.5 & 847.5\\ 
		Buildings, Zip Codes & 7.90~M & 235.1 & 185.7  \\ 
		Buildings, Countries & 112~M & 949.1 & 465.8 \\ \hline
	\end{tabular}
	\caption{Performance of polygon-polygon joins.}
	\label{tab:poly-poly}
\end{table}

\myparagraph{Analysis and Discussion.}
As we previously discussed, joins might process a single grid cell multiple
times. For example, a single grid cell of the Building data can intersect
multiple cells from the County data, and vice versa. In the case of the Building
\dataset, the polygons are significantly smaller compared to the data extent
(the entire world). Therefore, using a small grid cell size to accommodate these
polygons greatly increases the number of cells that must be
processed during the join. Thus, this situation represents a worst case scenario for
\spade, and we believe this presents an opportunity for query optimization
strategies in the future.
Also, similar to selection queries, \io dominates the execution times of joins as well.

\setlength{\columnsep}{5pt}%
\begin{wrapfigure}{r}{0.4\linewidth}
	\centering
	\includegraphics[width=\linewidth]{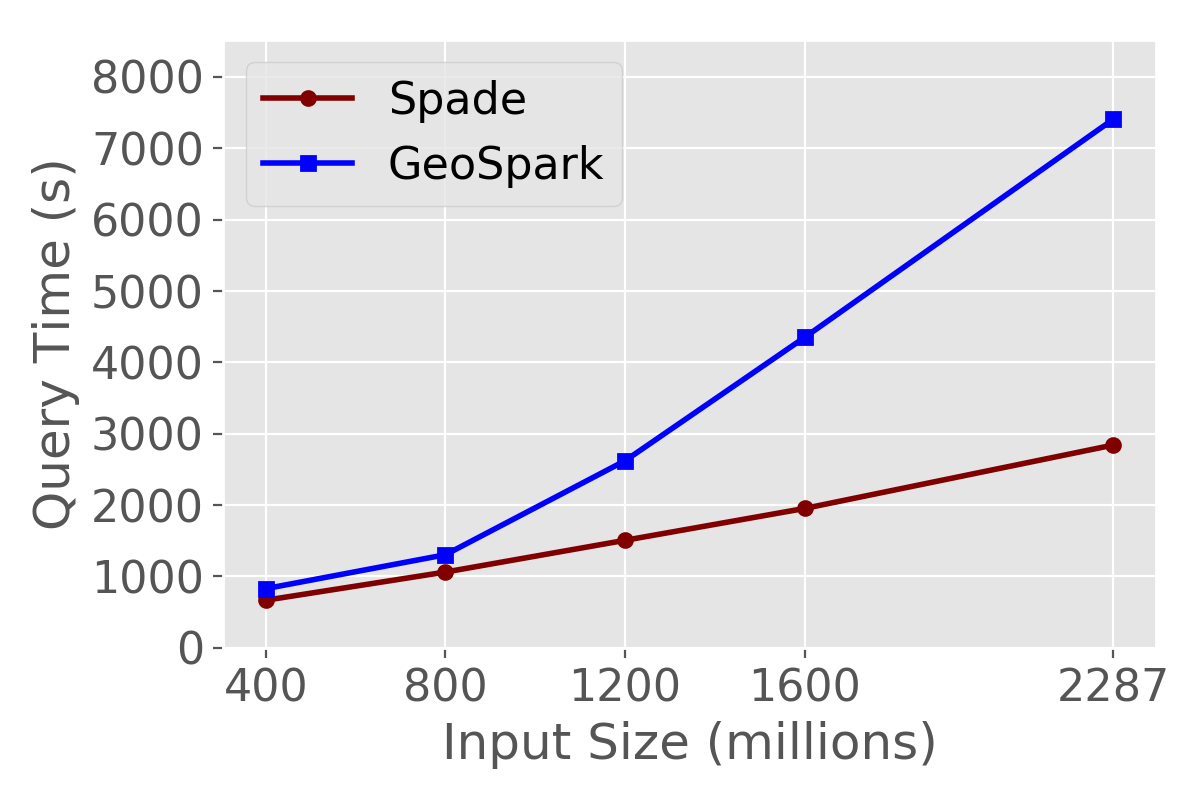}
	\vspace{-0.3in}
	\caption{Scaling with input size (Twitter-Zip Code join).}
	\label{plot:join-scalability}
	\vspace{-0.2in}
\end{wrapfigure}
In case of GeoSpark, we noticed that the running time of joins significantly increases once the data size (in terms of number of objects) becomes larger than a particular value. To confirm this, we executed the join query between subsets of Twitter \dataset with the Zip Code polygons. These subsets were extracted by increasing the time range of the tweets.
\fig{plot:join-scalability} shows the results from this
experiment. 
While the scaling is still linear, the slope however
increases once the number of points grows beyond one billion. Therefore, to obtain the best timings for GeoSpark in our evaluation, we split the Twitter data into multiple smaller subsets (approximately 600M points each), and run the join on each subset. This resulted in over 2X speedup in the total running time when compared to executing a single join with the entire Twitter data.
On the other hand, \spade scales better with increasing data sizes.

\subsection{Distance-Based Queries}
\label{sec:distance}

\begin{figure}[b]
	\centering
	\includegraphics[width=0.7\linewidth]{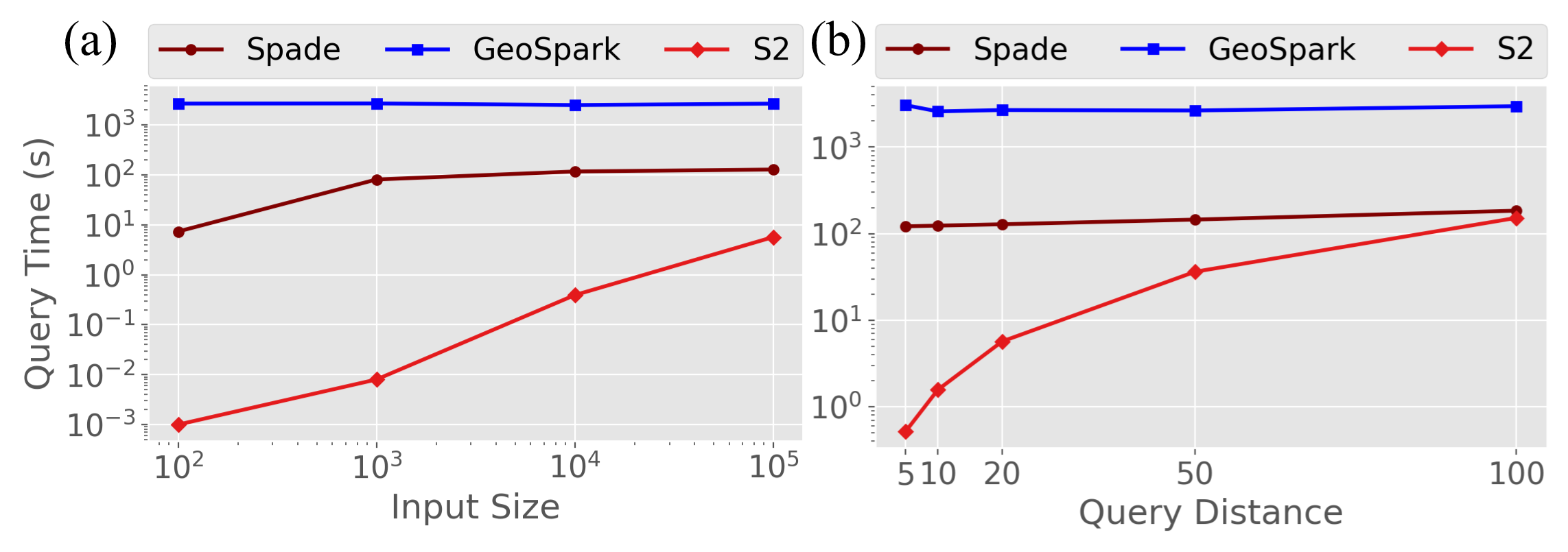}
	\vspace{-0.1in}
	\caption{Performance of distance-based joins.
		(a)~Varying number of points used with a constant query distance of 20~m.
		(b)~Varying query distance for a join with 100K points.}
	\label{plot:distance-join}
\end{figure}

We evaluated the performance of \spade for distance-based joins.
For this evaluation, we generated a random set of points within the spatial
extent of the Taxi data and joined it with the Taxi data. 
\fig{plot:distance-join}(a) compares the performance of \spade, GeoSpark, and S2 when the size
of the random points varies from 100 to 100K. We set the query distance to 20m for this experiment.
\fig{plot:distance-join}(b) compares the performance of the three approaches when the size of 
point set is fixed at 100K, and the query distance is varied between 5m and 100m.
S2 has the best performance for this class of queries, followed by \spade and GeoSpark respectively.
This is because the S2 point index is optimized for distance-based queries.

\myparagraph{Analysis and Discussion.}
In \fig{plot:distance-join}(a), we observe an interesting phenomenon with \spade: 
the query time for input size 100
is much faster than when the input size is 1000, after which the time
stabilizes.
This is because 
the number of intersections between the generated canvas polygons is much
lower for 100 points compared to the other sizes, thus requiring considerably less 
processing.
Otherwise, both \spade and GeoSpark have stable running times with varying
parameters. On the other hand, S2's time is dependent on the result size.

Note that, for this experiment, the input coordinates had to be converted into a meter-based
projection system. As discussed earlier, this is performed in the shaders on the fly by \spade 
during query execution.
While GeoSpark also supports this conversion, 
it took a significant amount of time ($>$ 2~hours).
We therefore decided to pre-convert the coordinates, and used this in the experiments.
In other words, the running times shown for GeoSpark do not reflect the time taken to convert the 
coordinates.

It is also interesting to note that running times using S2 increases with increasing query distance --
as the distance increases, the running time of S2 becomes closer to that of \spade.
This is due to the structure of the S2 index, where more index cells need to be covered for larger distances.

\subsection{\knn Queries}
\label{app:knn}
We use the same set of generated points that was used in the evaluation of distance-based queries for \knn queries 
over the Taxi data.

\setlength{\columnsep}{5pt}%
\begin{wrapfigure}{r}{0.35\linewidth}
	\centering
	\vspace{-0.2in}
	\includegraphics[width=\linewidth]{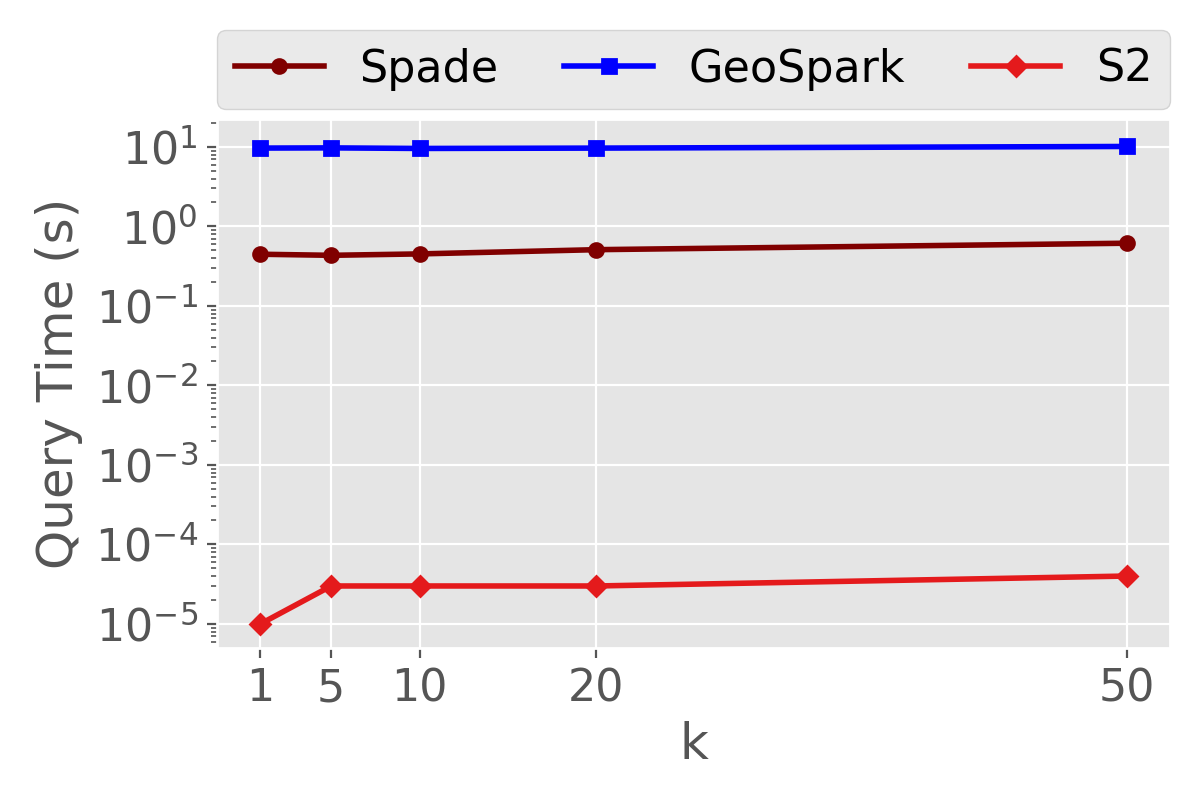}
	\vspace{-0.3in}
	\caption{Average time for 100 knn queries with varying k.}
	\label{plot:knn-select}
	\vspace{-0.25in}
\end{wrapfigure}
\myparagraph{kNN Selection.}
\fig{plot:knn-select} plots the average time taken for 100 \knn queries with the value of $k$
varying from 1 to 50. 
S2 takes just a few milliseconds
to execute 100 queries and is significantly faster than \spade and GeoSpark. 
This is again because S2's index is optimized to handle such queries.

\begin{figure}[h]
	\centering
	\includegraphics[width=0.7\linewidth]{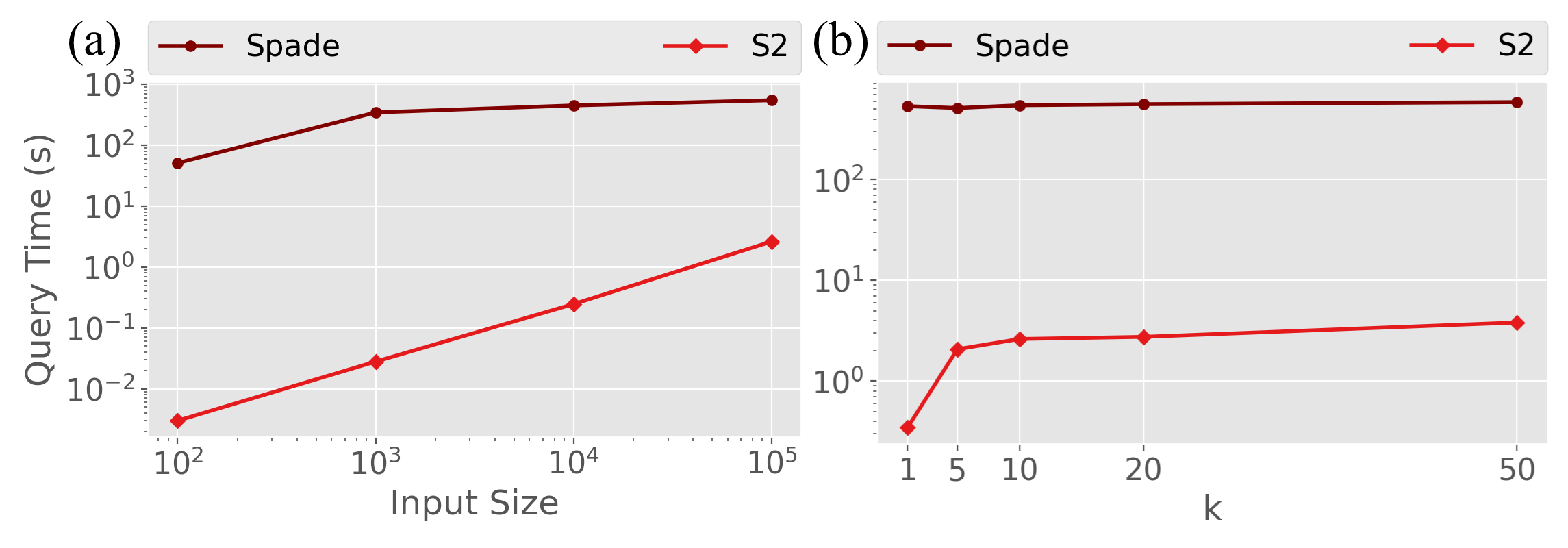}
	\vspace{-0.1in}
	\caption{kNN joins. 
		(a)~Varying k. Point size = 100k.
		(b)~Varying point size. k = 10.
	}
	\label{plot:knn-join}
	\vspace{-0.2in}
\end{figure}

\myparagraph{kNN Join.}
\figs{plot:knn-join}(a) and \ref{plot:knn-join}(b) show the performance
of \spade and S2 for varying sizes of the point data and different values of $k$, respectively.
Similar to kNN select, S2 performs better than \spade for this query class. 
Given this significant performance difference, 
it will be interesting to investigate whether S2's index can be adapted to be used with
the GPU operators as well.
Note that GeoSpark does not support \knn joins.

\input{syn-exp}

%% file: syn-exp.tex
\subsection{Synthetic Data}
\label{app:synthetic}

In this section we evaluate the performance of selection and join queries on synthetically generated data.
This evaluation allows us to test not only the scalability of \spade, but also its performance for different data distributions
in a controlled manner. 

\myparagraph{\Datasets and queries.}
The synthetic data was generated using the \textit{Spider spatial data generator}~\cite{spider}.
We generated four classes of \datasets comprising of points and rectangles. 
For each class, we generate 5 sizes as shown in Table~\ref{tab:synthetic}. Note that the number of rectangles is chosen to
have the same number of vertices as the point \datasets.
\begin{enumerate}[leftmargin=10pt]\denselist
	\item \textbf{Uniform Points}: A collection of points uniformly distributed over a unit square. 
	\item \textbf{Gaussian Points}: A collection of points normally distributed over a unit square. 
	\item \textbf{Uniform Boxes}: A collection of axis parallel rectangles of varying sizes uniformly distributed over a unit square. 
	\item \textbf{Gaussian Boxes}: A collection of axis parallel rectangles of varying sizes normally distributed over a unit square. 
\end{enumerate}

\begin{table}[t]
	\centering
	\footnotesize
	\begin{tabular}{|c||l|}\hline
		\textbf{Name} &  {\textbf{No. of Primitives (in millions)}} \\ \hline \hline
		Uniform Points & 40, 80, 120, 160, 200 \\ 
		Gaussian Points & 40, 80, 120, 160, 200 \\ 
		Uniform Boxes & 10, 20, 30, 40, 50 \\ 
		Gaussian Boxes & 10, 20, 30, 40, 50 \\ \hline
	\end{tabular}
	\caption{Synthetic \datasets used in our evaluation.}
	\label{tab:synthetic}
	\vspace{-0.2in}
\end{table}

For selection queries, we chose one of the polygons from the NYC Neighborhood data that was used in the Taxi experiments (\fig{plot:selection}(a)),  centered it
on the unit square, and scaled it to vary the extent of its bounding box in order to control the selectivity of the query.
In particular, we vary the range such that the width (and height) of the bounding box of the polygon varies from 0.1 to 0.5 (\ie covers half the width and height of the unit square)
in increments of 0.1.

For join queries, we generated five \textit{parcel} \datasets having 1000, 2500, 5000, 7500, and 10000 parcels (non-intersecting rectangles of varying sizes) respectively.

\begin{figure}[!b]
	\centering
	\includegraphics[width=0.7\linewidth]{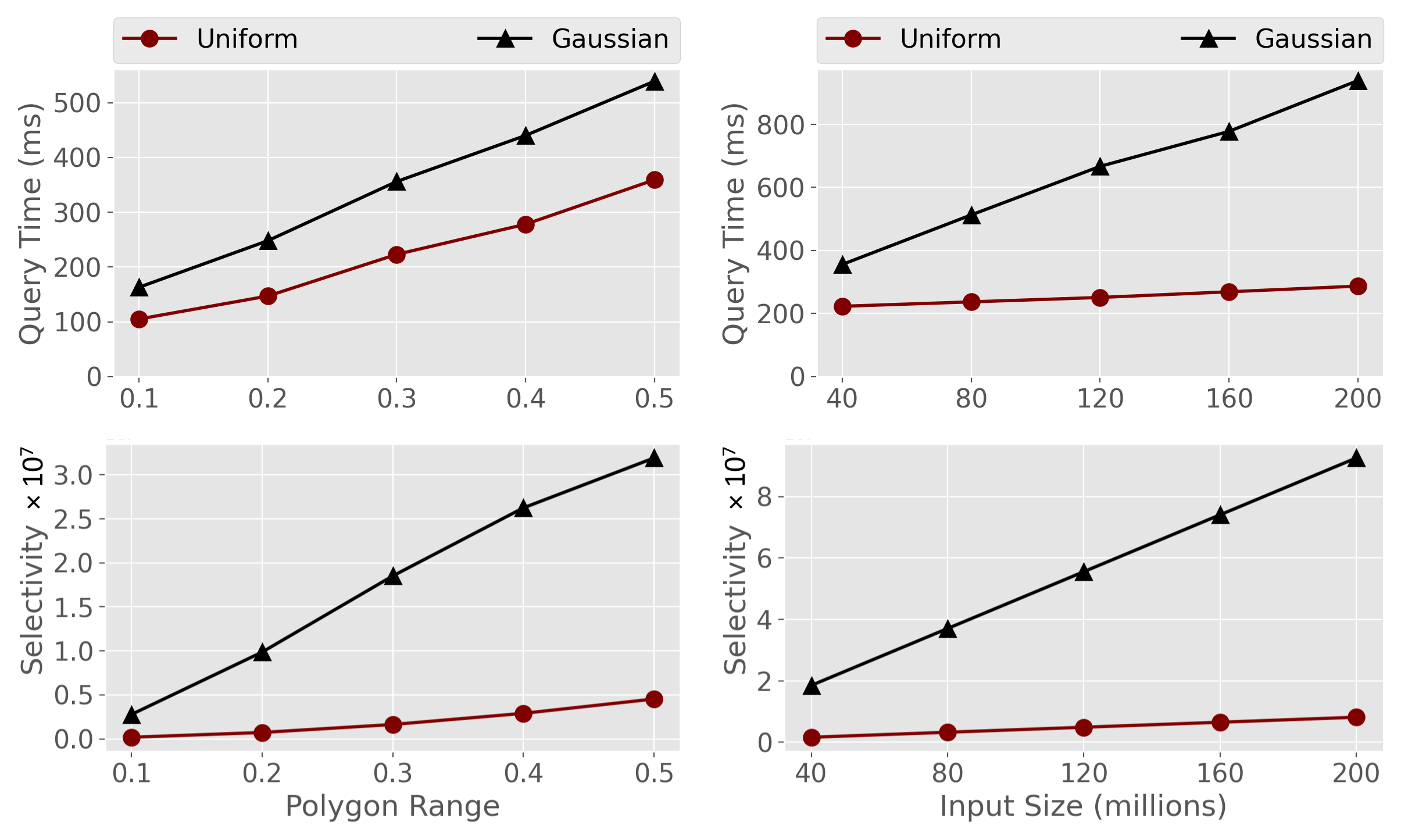}
	\caption{Selection over uniform and Gaussian point data. 
		\textbf{(Left Top)}~Varying the extent of the query polygon on the 40~million points \dataset.
		\textbf{(Right Top)}~Varying the input data size. The range of the query polygon is fixed at 0.3.
		\textbf{(Bottom)}~The selectivity of the queries corresponding to the top row.
	}
	\label{fig:syn-pts-sel}
	\vspace{-0.1in}
\end{figure}

\myparagraph{Performance.}
\fig{fig:syn-pts-sel} (left top) shows the running times of selection queries over the point \datasets with varying selectivity.
The running times are proportional to the selectivity. Queries over the Gaussian data takes more time than those over uniform data 
since they have a higher selectivity (\fig{fig:syn-pts-sel}
(bottom)). 
We would like to note that the index for both the uniform and Gaussian points were created 
using the same parameters. Thus, the cells of the grid index filtered in the index
filtering stage is the same for a given query polygon for both the \datasets. 
However, since the number of points within
these cells is significantly more for the Gaussian data (more points are concentrated at the center of the unit square which is also the center of the query polygon), 
the data transfer time becomes higher. 
Thus, the above noticed time increase is mainly dominated by the \io time, while the processing time on the GPU varies only by a few milliseconds.

\begin{figure}[t]
	\centering
	\includegraphics[width=0.7\linewidth]{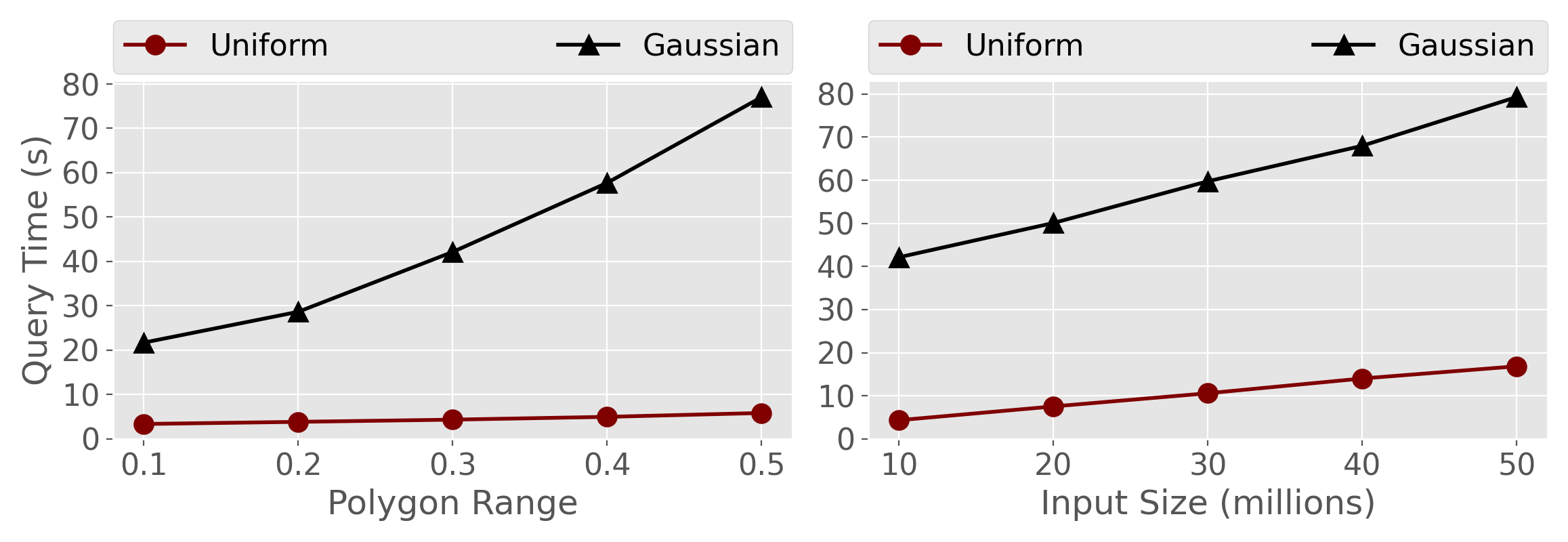}
	\caption{Selection over uniform and Gaussian box data. 
		\textbf{(Left)}~Varying the extent of the query polygon on the 10~million boxes \dataset.
		\textbf{(Right)}~Varying the input data size. The range of the query polygon is fixed at 0.3.
	}
	\label{fig:syn-box-sel}
\end{figure}

\fig{fig:syn-pts-sel} (right top) shows the running time when the query polygon is fixed and the input size is varied. 
As input size increases, the 
result size increases significantly for the Gaussian data when compared to the uniform data. This is reflected in the
running times as well.
\fig{fig:syn-box-sel} shows the performance of selections over polygonal \datasets with varying selectivity (left) and varying input sizes (right).
We see a trend similar to that of the point data.

Unlike the selection experiments on real data, we can control
different aspects of the synthetic data. 
In particular, we can fix the complexity
of the query polygon. Thus, we can see a clear trend of the running times scaling linearly with the query selectivity.
On the other hand, it is difficult to see such trends using real data, especially in a spatial setting. This is because
even when two query polygons have the same bounding box, they can have different selectivity due to their shape and complexity (\eg a star query constraint vs. a square constraint, both with the same bounding box). 

\begin{figure}[b]
	\centering
	\includegraphics[width=0.7\linewidth]{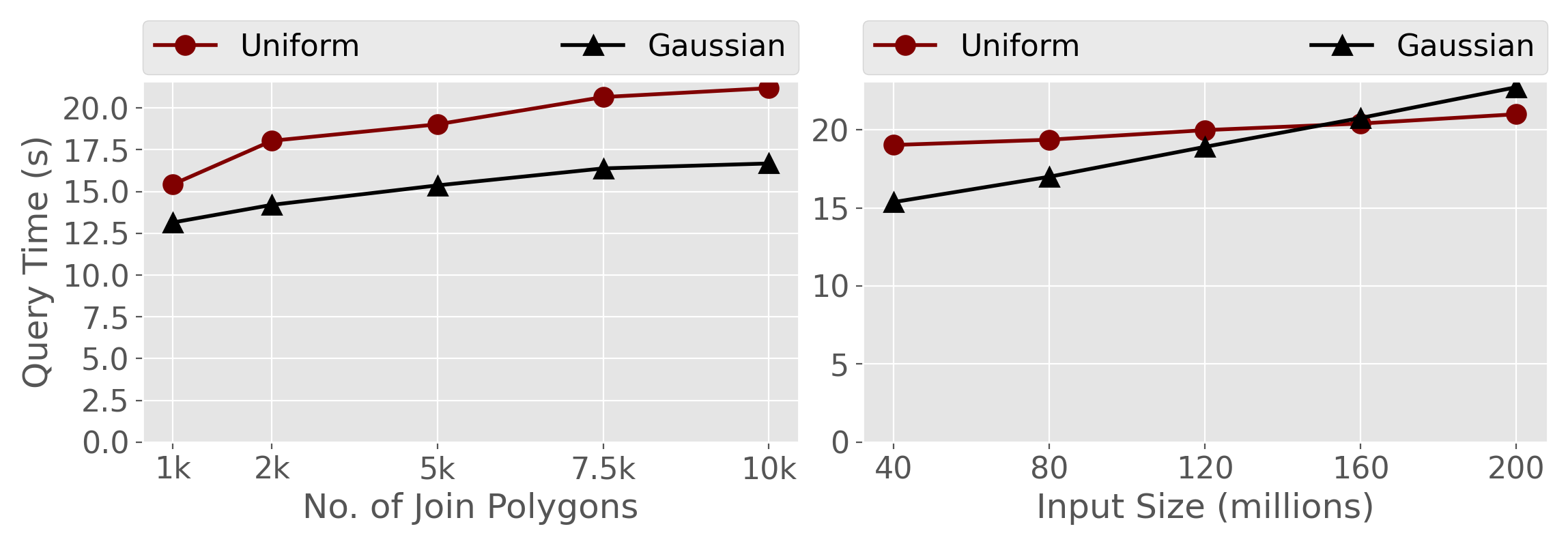}
	\caption{Performance of Point-Polygon joins. 
		\textbf{(Left)}~Varying the number of parcels in the join with the 40~million point \datasets.
		\textbf{(Right)}~Varying the size of the point data. Polygons with 5000 parcels was used for this join.
	}
	\label{fig:syn-pts-join}
\end{figure}

\fig{fig:syn-pts-join} shows the performance of point-polygon joins. Unlike the selection queries, 
the running time of joins over the Gaussian data is smaller than that over the uniform data. 
Recall that to execute joins, the filter phase first identifies all pairs of index grid cells that are then joined in a loop. 
Unlike the selection query, the join query used for this experiment is designed to cover the entire unit square. 
This requires the entire data to be transferred at least once to the GPU. 
However, since most of the points in the Gaussian data are concentrated at the center of the unit square, 
most of the query results are focused on a smaller set of grid cell pairs. 
Thus, larger amounts of contiguous data gets transferred into the GPU in a smaller set of rendering passes during the out-of-core processing.
As a side-effect of this, when the GPU is processing (or rendering) a subset of the data, data that is to be processed next during the same (but larger) rendering pass 
gets transferred simultaneously in a pipelined fashion. 
On the other hand, in the join over uniform data, data transfers are
spread uniformly across all the rendering passes, which we believe
reduces the ``pipeline" effect since the 
data transfer for the next rendering pass has to wait until the current rendering is complete. We believe this to be the primary reason for the shown trend where joins over Gaussian data is faster than that over uniform data.
As input size increases, we however notice that this difference becomes smaller.

\begin{figure}[t]
	\centering
	\includegraphics[width=0.7\linewidth]{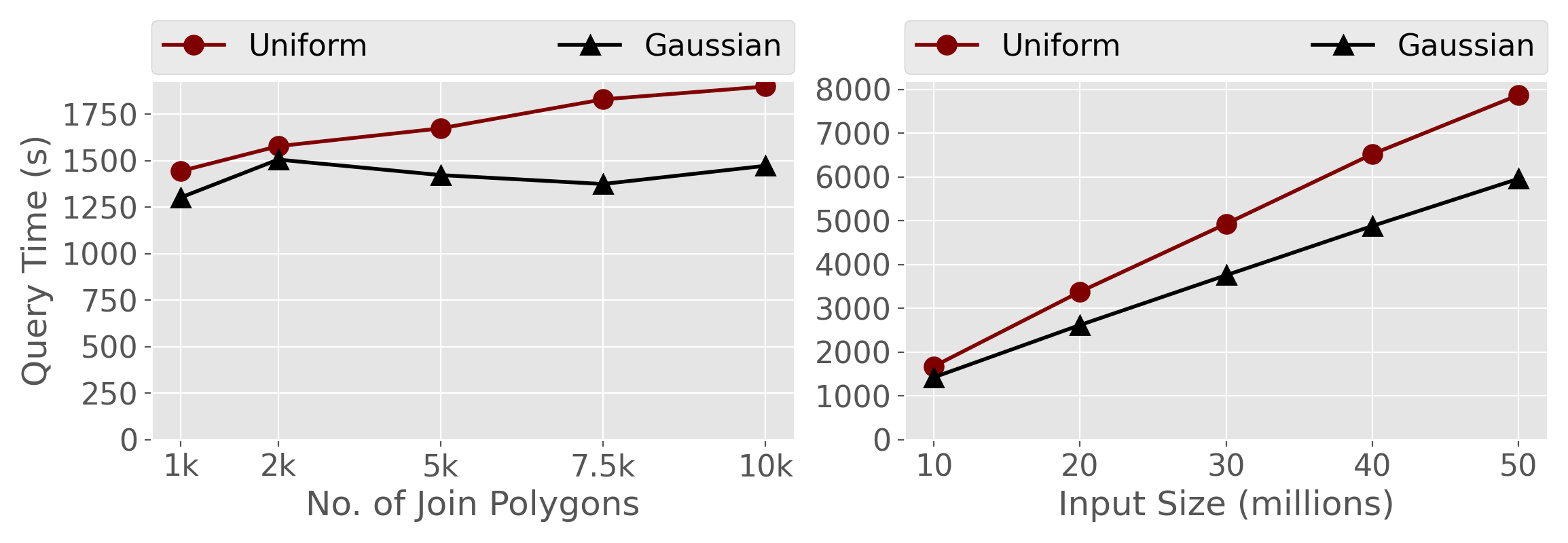}
	\caption{Performance of Polygon-Polygon joins. 
		\textbf{(Left)}~Varying the number of parcels in the join with the 10~million box \datasets.
		\textbf{(Right)}~Varying the size of the box data. Polygons with 5000 parcels was used for this join.
	}
	\label{fig:syn-box-join}
\end{figure}

\fig{fig:syn-box-join} shows the performance of polygon-polygon joins. We note that the above trend, where queries over Gaussian data
are faster than those over uniform data, applies for this case as well.

%% file: discussion.tex
\section{Limitations and Discussions}
\label{sec:discussions}

\myparagraph{Catering to the Long Tail.} 
\spade is not a replacement to libraries such as S2 and systems such
as GeoSpark -- they can attain better performance and
scalability in different scenarios.
For example, since S2 is optimized for distance based queries, it can
be faster than \spade for some queries. However, it requires large
main memory hardware to run, while \spade can handle these queries on
(commodity) systems with limited memory, albeit requiring a longer
running time.
Similarly, GeoSpark due to its distributed nature is better suited for 
efficiently executing batch-based queries over very large \datasets
which \spade may not be able to support.
However, given that analysts making up the long tail often use
commodity hardware for their processing, we believe that \spade would
provide them with a viable alternative and thus democratize large
scale spatial analytics.

\myparagraph{Containment Queries.}
\spade currently supports intersection constraints for queries 
over line and polygonal \datasets. 
Support for containment constraints can be added by simply
treating each line (or polygon) as a collection of
vertices, and testing for containment of this collection, \ie point-based queries
can be reused for this purpose.
Note, however, that for points an intersection is equivalent to a containment test.

\myparagraph{CPU-GPU hybrid execution.}
The currently supported queries (which are purely spatial) are such that it can be decomposed 
into a set of intersection tests and distance computations. This is not only amenable to parallelization, 
but can also efficiently make use of the graphics pipeline using the spatial algebra. 
Thus, we implemented these queries purely using the GPU. 
However, we believe that a hybrid approach will be beneficial when more complex queries are involved, 
possibly composed of both relational and spatial constraints. This also opens up new
research problems to support dynamic scheduling and realization of query plans.

\myparagraph{Improving I/O Performance.}
Even though \io is a bottleneck in \spade's execution times,
the CPU-GPU bandwidth is continuously increasing through the introduction fast
interconnects~\cite{Clemens2020}.  Therefore, advances in the hardware will 
naturally improve the performance of \spade.
In addition to relying on technological advances, other strategies could also
be used to further improve the \io performance. 
For example, a pipeline-based approach could 
benefit GPUs with larger memory (such as desktop GPUs),
wherein, while current batch is being executed in the GPU, 
the data from the next batch can be simultaneously loaded.
Since CPU memory is in general 
much larger than the GPU memory, \io can 
also be optimized by 
incorporating an appropriate caching policy
that takes into account the order of operations.
Recall that in spite of the memory bottlenecks, \spade already displayed very good performance.
Therefore, we left these optimizations to be implemented in the next version of \spade.

\myparagraph{Indexing Strategies and Tuning.}
\spade currently uses a grid-based indexing strategy,
with the size of the grid cell being specified by the user. 
For our experiments, we used a general rule of thumb 
based on the GPU memory size for this purpose.
Since, we focused on processing as much data as possible on the GPU (to maintain high
GPU occupancy), this inadvertently also increased the amount of data
filtered by the index.
It will be interesting to explore automatic tuning based on
system configuration and/or query workload.

Alternatively, other indexing strategies can also be used in a similar fashion.
For example, if an R-tree based strategy is used, the bounding
polygons of the R-tree leaves can be stored where the leaf sizes are
adjusted similar to the grid cell size above.
Note that, the index filtering, however, need not use the traditional R-tree traversal,
and instead simply perform selections/joins on the bounding polygons.

\myparagraph{Multi-GPU Support.}
While OpenGL provides an easy high level API
for the graphics pipeline, its multi-GPU support 
is limited except in the case of Nvidia hardware and only under specific conditions
(using an SLI interface or on Linux). 
In the next iteration of \spade, we intend to switch to use the recent 
Vulkan~API~\cite{vulkan} (which can be seen as an extension of OpenGL)
that provides a much more fine-grained control over GPUs.
This would also allow diverse GPUs (such as a discrete and an integrated GPU) to be used together,
further improving performance even on laptops.

%% file: related.tex
\section{Related Work}
\label{sec:related-work}

The advent of hardware with multiple processing units has
led to the design of new approaches that use them for spatial query
processing. In particular, GPUs and multi-node clusters 
are extremely popular for this purpose.

Doraiswamy~et~al.~\cite{STIG2016} proposed a kd-tree based index
optimized for GPUs to support interactive selections over 
point data. 
Zhang~et~al.~\cite{Zhang2012a, Zhang2012d} used GPUs to join points
with polygons. They index the points with a Quadtree to achieve load
balancing and enable batch processing.
They then extended their spatial join framework~\cite{Zhang2015} 
to handle larger point \datasets. 
However, to improve efficiency, they truncate coordinates to 16-bit integers, 
thus resulting in approximate joins. 
Tzirita Zacharatou~et~al.~\cite{rasterjoin} proposed a rasterization-based
approach to compute spatial aggregations between a set of points and polygons.
Aghajarian~et~al.~\cite{gcmf} proposed a GPU approach for both 
point-polygon and polygon-polygon joins. However, their 
approach only works on data that fits in GPU memory.
Sainju~et~al.~\cite{Sainju2020} proposed a grid-based join
approach with the intent of mining co-located spatial events.
Wang and others~\cite{Wang2012,Gao2018,Baig2020} proposed 
an approximate GPU-based join approach.
There have also been GPU-based approaches for heatmap queries~\cite{Yong2015},
creating and querying R-trees~\cite{Prasad2015}, and improving the
filtering phase of spatial queries~\cite{Sun2003,Aghajarian2017}.
%
All the above approaches use CUDA (except \cite{Sun2003} and \cite{rasterjoin}), and are hence 
restricted to work only on Nvidia GPUs.
While we reviewed a representative sample of GPU-based approaches,
to the best of our knowledge, existing GPU-based spatial techniques are specific 
to a given query type.

OmniSci~\cite{omniscidb} provides a complete GPU-based 
SQL engine that also supports certain spatial queries. 
Similar to the above approaches, OmniSci is tied to Nvidia GPUs.
Additionally, it is an in-memory system, and thus cannot be used
to handle large \datasets on commodity desktops and laptops.

An initial prototype for the algebra and canvas model was described in
\cite{gpu-algebra}.  The prototype was developed as a proof of concept
and it only supports spatial selection of points over data that fits in memory.
\spade, in contrast, is a full-fledged query processing engine. As
discussed above, we had to address several challenges to support the
efficient evaluation of a rich set of spatial queries on commodity
hardware.

Recently, several spatial database systems have been proposed using MapReduce
such as Hadoop-GIS~\cite{hadoop-gis}, Simba~\cite{Xie:2016:SEI:2882903.2915237} and GeoSpark~\cite{geospark}.
Eldawy and Mokbel~\cite{Eldawy2016} provide a comprehensive survey of these
approaches. 
Pandey~et~al.~\cite{pandey@vldb2018} presented a
detailed experimental evaluation of the state-of-the-art 
cluster-based spatial analytics systems.
Pandey~et~al.~\cite{pandey@dsc2020} also evaluated common 
spatial libraries. 
As we discussed, these approaches are effective but out of reach for
many analysts.

%% file: conc.tex
\section{Conclusions}
\label{sec:conclusions}
  
In this paper we presented \spade, a query engine that implements the
recently proposed GPU-friendly spatial model and algebra~\cite{gpu-algebra} to support
spatial queries over large \datasets.  
\spade introduced two canvas-specific indexes
to improve efficiency of spatial queries and a query optimizer to
plan and execute queries on the GPU.
We experimentally compared \spade with other approaches using 
large, real-world \datasets. The results show that queries evaluated by
\spade on a laptop with an Nvidia GPU  have running times that are smaller or
comparable to those of the cluster-based GeoSpark on a 17-node cluster, and of
S2 running on a server with large memory. These results suggest that
\spade has the potential to make large-scale spatial analytics within reach for
a broad set of stakeholders.
Since \spade was designed to use the relational schema to store spatial
\datasets, indexes, as well as the necessary meta-data, it can easily 
be integrated into existing relational database systems.

The novel approach to spatial query processing used by \spade not only opens
new research opportunities in spatial query optimization and indexing,
but also enables complex queries that were not possible before--e.g.,
accurate distance-based queries with respect to complex objects
that are not supported in existing systems due to its complexity
can be accomplished in \spade with a minimal overhead.
Furthermore, with the recent Spark~3.x adding support for GPUs,
we believe a significant performance boost can be obtained by integrating 
\spade with systems such as GeoSpark.